\documentstyle[12pt,epsf]{article}
\newcommand{\bb}{\begin{equation}}   
\newcommand{\ee}{\end{equation}}
\newcommand{\beqa}{\begin{eqnarray}} 
\newcommand{\eeqa}{\end{eqnarray}}

\def\beq{\begin{equation}}   \def\eeq{\end{equation}}

\newcommand{\gsim}{\lower.7ex\hbox{$
\;\stackrel{\textstyle>}{\sim}\;$}}
\newcommand{\lsim}{\lower.7ex\hbox{$
\;\stackrel{\textstyle<}{\sim}\;$}}

\newcommand{\matel}[3]{\langle #1|#2|#3\rangle}

\newcommand{\ra}{\rightarrow}

\newcommand{\Lam}{\Lambda_{QCD}}
\newcommand{\re}[1]{Ref.~\cite{#1}}

\newcommand{\prt}{perturbative }

\newcommand{\eq}[1]{Eq.\hspace*{.1em}(\ref{#1})}
\newcommand{\eqs}[1]{Eqs.\hspace*{.1em}(\ref{#1})}
\newcommand{\Eq}[1]{Eq.\hspace*{.1em}(\ref{#1})}

\renewcommand{\Im}{\mbox {Im}\:}

\newcommand{\be}{\beta}
\newcommand{\ga}{\gamma}
\newcommand{\de}{\delta}
\newcommand{\al}{\alpha}
\newcommand{\as}{\alpha_s}

\newcommand{\GeV}{\,\mbox{GeV}}

\begin{document}
\begin{titlepage}
\renewcommand{\thefootnote}{\fnsymbol{footnote}}

\begin{center} \Large
{\bf Theoretical Physics Institute}\\
{\bf University of Minnesota}
\end{center}
\begin{flushright}
TPI-MINN-96/05-T\\
UMN-TH-1426-96\\
CERN-TH/96-113\\
hep-ph/9605465\\
\end{flushright}
\vspace{.3cm}
\begin{center}
{ \Large
{\bf Operator Product Expansion, Heavy 
Quarks, QCD Duality and its Violations}} 
   
\end{center}
\vspace*{.3cm}
\begin{center} {\Large Boris Chibisov, R. David Dikeman, 
M. Shifman} \\
\vspace{0.4cm}
{\it  Theoretical Physics Institute, Univ. of Minnesota,
Minneapolis, MN 55455}\\
\vspace{0.4cm}
and \\
\vspace*{.4cm}
{\Large N.G. Uraltsev}\\
\vspace{0.4cm}
{\it TH Division, CERN, CH 1211 Geneva 23, Switzerland}\\
{\small and}\\
{\it St.Petersburg Nuclear Physics Institute,
Gatchina, St.Petersburg 188350, Russia} 

\end{center}

\vspace*{.2cm}
\begin{abstract} 

The quark (gluon)-hadron duality constitutes a basis for the
theoretical treatment of a wide range of inclusive processes --
from hadronic $\tau$ decays and $R_{\rm e^+e^-}$, to semileptonic 
and nonleptonic decay rates of heavy flavor hadrons. A theoretical 
analysis of these processes is carried out by using the operator 
product expansion (OPE) in the Euclidean domain, with
subsequent analytic continuation to the Minkowski domain.
We formulate the notion of  the quark (gluon)-hadron duality
in quantitative terms, then classify various contributions leading to 
violations of duality. A prominent role in the violations 
of duality seems to belong to the so called exponential terms which, 
conceptually, may represent the (truncated) tail of the power series.
A qualitative model, relying on an instanton background field,
is developed, allowing one to get an estimate
of the exponential terms.  We then discuss a number of 
applications,
mostly from heavy quark physics. 
  
\end{abstract}

\vfill
\noindent
CERN-TH/96-113\\
April 1996
\end{titlepage}

\section{Introduction}

Nonperturbative effects have been analyzed in QCD in the framework
of the operator product expansion (OPE) \cite{OPE,OPE1} since the 
inception of QCD. Recently a remarkable progress has been achieved 
along these lines in the heavy quark theory (for reviews see Ref. 
\cite{HQ}).
A large number of applications of OPE in heavy quark theory 
refer to quantities of the essentially Minkowski nature, e.g. 
calculations of the inclusive 
decay widths, spectra and so on. Wilson's operator product expansion
{\em per se} is formulated in the Euclidean domain. The expansion
built in the Euclidean domain and by necessity truncated, is
translated in the language of the observables through an analytic 
continuation. An indispensable element of this procedure,
the so called quark-hadron duality, is always assumed, most often  
tacitly. This paper is devoted to the discussion of  
quark-hadron duality and deviations from it. Although the issue will 
be considered primarily in the context of heavy quark theory,
the problems we will deal with are quite general and are by no 
means confined to heavy quark theory.  For instance, 
determination of $\alpha_s$ from the hadronic width of $\tau$ --
a problem of paramount importance now under intense 
scrutiny \cite{tau} -- falls into this category. It is known
\cite{Shif} that deviations from duality may be conceptually related 
to the behavior of the operators of high dimension in OPE. 
Unfortunately, very little is known about this behavior in the 
quantitative sense, beyond the fact that the expansion is asymptotic
\cite{Shif}. Therefore we are forced to approach the problem from 
the other side -- engineering {\em  a  model} which allows us to start 
discussing deviations from duality. 
The model is based on instantons, but by no means is derivable 
in QCD. Moreover, it does not exhaust all mechanisms which might 
lead, in principle, to deviations from duality, focusing, rather, on one 
specific contribution -- the so called exponential terms. Nevertheless, 
it seems to be physically motivated  and can serve for 
qualitative analysis at present,  and  as a guideline for future 
refinements. 

Indeed, the (fixed size) instanton contribution to correlation 
functions with large momentum transfers can be interpreted
as a mechanism in which the large external momentum is 
transmitted through a soft coherent field configuration. Speaking 
graphically, the large external momentum is shared by a very large 
number of quanta so that each quantum is still relatively soft. 
It is clear that this mechanism is not represented in the practical 
version of OPE \cite{OPE1}, and, thus, gives an idea of how strong 
deviations from duality might be.

One of the most interesting aspects  revealed in this model
is the distinct nature of exponential contributions absent in  
practical 
OPE, both in the Euclidean and the Minkowski domains. 
If in the former the 
exponential effects die off  fast enough, in the latter,
deviations from duality are suppressed to a lesser extent -- the 
exponential fall off is milder, and it is modulated by oscillations. 
These features seem to be so general that most certainly they will 
survive in future treatments which, hopefully, will  be significantly 
closer to  fundamental QCD than our present consideration. 

If one accepts this model, at least for orientation,  many interesting 
technical problems arise.  Instanton contributions in  
heavy quark theory were previously discussed more than once 
\cite{GO,Chy}. 
Although the corresponding analysis seemed 
rather straightforward at first sight, it resulted  in some apparent 
paradoxes; for instance, the instanton contribution to the spectrum of 
the inclusive heavy quark decays seemingly turned out to be 
parametrically larger 
than the very same contribution to the total decay rate 
\cite{Chy,AFalk}.  The puzzle is readily solvable, however: one 
observes that the problem lies in 
the separation of the exponentially small terms from the 
``background" of the power terms of OPE; this is a subtle
and, generally speaking, ambiguous procedure,  particularly in the 
Minkowski domain, and depends on the  
specific quantity under consideration. We will dwell on this issue at 
length in the present paper.

We begin, however, with a brief formulation of the very notion of 
duality
(Sect. 2).
Quantifying this notion is an important task by itself. In modeling 
deviations from duality, the adoption of the following attitude is 
made 
so as to stay on safe ground: we will
try to develop a model yielding a conservative  estimate on the {\em 
upper bound}
for deviations from duality. In other words, given the prediction for 
this or that quantity based on duality (i.e.  spectra, total inclusive 
widths and so on),
one establishes the accuracy with which this prediction is expected to 
be valid.  In this way one sets the lower  
limit on the energy release needed
to achieve the required accuracy. For this limited purpose even a 
crude model, such as the instanton model to be discussed below, may 
be sufficient, perhaps, after some minor refinements.

Why is this attitude logical? If we knew in detail some specific 
mechanism omitted in the theoretical calculations -- whether  
associated with the truncation of practical OPE, or  due to  other 
sources 
-- we could include it in the theoretical prediction for the cross 
sections and say that the actual hadronic cross section is dual to this 
new improved prediction. Thus, paradoxically, the very nature of 
duality 
implies that deviations from it are always estimated  roughly.
Analyzing deviations from  duality at each given stage of
development of QCD is  equivalent to analyzing our ignorance, rather 
than our knowledge. At the present stage, as was already mentioned,
our knowledge is, more or less, limited to practical OPE.  

A much more ambitious goal is developing a framework  suitable 
for 
actual {\em calculation} of extra contributions not seen in  
practical OPE. Although  the  instanton model is sometimes used for 
this purpose as well, one should clearly realize that quantitatively 
reliable results are not expected to emerge in this way.
This is a speculative procedure intended only for  qualitative 
orientation. We will occasionally resort to it only due to the 
absence of better 
ideas. One may hope that a universal qualitative  picture will be 
revealed {\it en route}, which will be robust enough to survive 
future 
developments of the issue. 

Although this problem -- estimates of deviations from duality --  is 
obviously of paramount practical importance, surprisingly little has 
been
said about this subject in the literature. Apart from some general 
remarks presented in Ref. \cite{Shif}, an attempt to discuss the 
issue in  a different (exclusive) context was made in \cite{BM}.  

Our paper is organized below as follows:
In Sect. 2 we outline the general principles behind duality and its 
violation.
Sect. 3 is devoted to general features of the exponential terms 
believed to be 
responsible
for duality violation. In  
particular, the distinction between their patterns in the Euclidean 
and Minkowski domains is explained here. 
In Sect. 4 we outline the instanton model we use as a framework 
to generate exponential terms. 
Sect. 5 illustrates our main points in  
what is, probably,  the most transparent example: $\rm e^+e^-$ 
annihilation and 
the hadronic 
decays of the $\tau$ lepton. In Sect. 6 we discuss the general 
features
of heavy quark 
decays in the instanton background.  In Sect.~7, we begin the
business of actual calculation -- to 
warm up, we consider a toy model where the  spins of all relevant 
fields are discarded to avoid technicalities. Section 8 is 
devoted to  actual heavy quark decays in QCD.  
The exponential contributions are 
estimated, both in the spectra, and in the inclusive decay rates, for 
the 
transitions of the heavy quark into a massless one. In section 9
we address the applied, but practically important, problem of 
deviations from duality in the semileptonic decays of $D$ and $B$ 
mesons. There are good phenomenological reasons  to believe that in 
 $D$ decays
these deviations are significant, of order 0.5. Adjusting parameters of 
the model in such a way as to explain these deviations, 
 we conclude that deviations from duality in the 
$B$ decays are expected to be negligibly small (in the total 
semileptonic decay rate and in the similar radiative processes). The 
effect seems to be 
 larger -- perhaps  even detectable -- in the inclusive 
nonleptonic rates.
The drawbacks and deficiencies of the model we use for the
estimates of the exponential terms are summarized in Sect. 10.
We present some comments on the vast literature treating the 
processes under discussion in Sect. 11.
Section 12 summarizes our results and outlines problems for 
future exploration. 

\section {Duality and the OPE}

Wilson's OPE is the basis of virtually all calculations of 
nonperturbative effects in analytical QCD. Since 
the very definition of duality relies heavily on Wilson's OPE, we first 
briefly 
review its main elements. For the sake of   definiteness, we will 
speak 
of the heavy quark expansion, although one should keep in mind
that the procedure is quite general; in other processes (e.g. the 
hadronic $\tau$ decays) the wording must be somewhat changed,
but the essence remains intact.

The original QCD Lagrangian is formulated at very short distances. 
Starting 
from this Lagrangian, 
one evolves it down, integrating out all fluctuations with frequencies
$\mu <\omega < M_0$ where $M_0$ is the original normalization 
point, and $\mu$ will be treated, for the time being, as a current
parameter. In this way we get the Lagrangian which has the
form
\bb
{\cal L} = 
\sum_n  C_n(M_0;\mu ){\cal O}_n(\mu )\, .
\label{10}
\end{equation}
The coefficient functions, $C_n$, represent the contribution of virtual 
momenta from $\mu $ to $M_0$. The operators, ${\cal O}_n$, enjoy 
the full rights of Heisenberg operators with respect to all field 
fluctuations with frequencies less than $\mu$. The sum in Eq. 
(\ref{10}) is infinite --
it runs over all possible Lorentz singlet gauge invariant operators
which possess the appropriate quantum numbers. The operators can 
be 
ordered according to their dimension; moreover, we can use the 
equations
of motion, stemming from the original QCD Lagrangian, to get rid  of 
some of the
operators in the sum. Those operators that are reducible to full 
derivatives give vanishing contributions to the physical (on mass 
shell) matrix elements, and can thus be  discarded as well.

Speaking abstractly, one is free
to take any value of $\mu$ in Eq. (\ref{10});
in particular, $\mu=0$ would mean
that {\em everything} is calculated and we have the full $S$ matrix,
all conceivable amplitudes, at our disposal.  Nothing is left to be done.
In this case Eq. (\ref{10}) is just a sum of all possible amplitudes.
This sum  then must be written in terms of physical hadronic 
states,
of course, not in terms of the quark and gluon operators
since the latter degrees of freedom simply do not survive scales 
below
some $\mu_{\rm had} \sim \Lambda_{\rm QCD}$.  

Needless to say,  present-day QCD does not allow the explicit 
evolution down to $\mu = 0$. 
Calculating the coefficient functions we have to stop somewhere,
at such virtualities that the quark and gluon degrees of freedom are
still relevant, and the coefficient functions $C_n(M_0, \mu )$ are still 
explicitly calculable.  On the other hand, for obvious reasons, it is 
highly desirable to have $\mu$ as low as possible.  In the heavy 
quark theory there is an additional requirement that $\mu$ must be 
much less than $m_Q$.  

 Let us assume that
$\mu$ is large enough so that $\alpha_s (\mu )/\pi \ll 1$
on the one hand, and small enough so that there is no large gap
between $\Lambda_{\rm QCD}$ and $\mu$. The possibility to make 
such a choice of $\mu$ could not be anticipated {\em a priori}
and is an extremely fortunate feature of QCD.
Quarks and gluons with  offshellness larger than $\mu$
chosen in this way will be referred to as  hard. 

 All observable amplitudes must be $\mu$ independent, of course.
The $\mu$ dependence of the coefficient functions $C_n$ must 
conspire with that of the matrix elements of the operators ${\cal 
O}_n$ in such a way as to ensure this $\mu$ independence of the 
physical amplitudes. 

What can be said about the calculation of the coefficients $C_n$?
Since $\mu$ is sufficiently large, the main contribution
comes from  perturbation theory. We just draw all relevant Feynman 
graphs and calculate them, generating an expansion in $\alpha_s 
(\mu )$
\begin{equation}
C_n= \sum_l a_l \alpha_s^l (\mu )\, .
\label{asum}
\end{equation} 
 (Sometimes some graphs will contain not only the  powers of 
$\alpha_s 
(\mu )$ but also  powers of $\alpha_s\ln (m_Q /\mu )$. This 
happens if 
the anomalous dimension of the operator ${\cal O}_n$ is 
nonvanishing,  or if a part of a
contribution to $C_n$ comes from characteristic momenta of order 
$m_Q$ and is, thus, expressible in terms of $\alpha_s (m_Q)$,
and we rewrite it in terms of $\alpha_s (\mu )$.) 

As a matter of fact expression (\ref{asum})
is not quite accurate theoretically. One should not forget that, 
in doing the loop integrations in $C_n$, we {\em must} discard the 
domain
of virtual momenta below $\mu$, by definition of $C_n(\mu )$. 
Subtracting  
this domain from the perturbative loop integrals, we introduce in 
$C_n$
power corrections of the type $(\mu /m_Q)^n$ by hand. 
In principle, one should recognize the existence of such corrections
and deal with them. The fact that they are 
actually present
was realized long ago \cite{OPE1}. Neglecting them, at the theoretical 
level, results in countless paradoxes which still surface from time to 
time in the literature.
If it is possible
to 
choose $\mu$ sufficiently small, these corrections may be 
insignificant
numerically, and can be omitted. This is what is actually done in 
practice. This is one of the elements of a simplification of the
Wilson operator product expansion.  The simplified version is 
called the {\em practical version of the OPE}, or practical OPE \cite{OPE1}. 

Even if perturbation theory may dominate the coefficient functions, 
they still also contain nonperturbative terms coming from short 
distances. 
Sometimes  they  are referred to as noncondensate nonperturbative 
terms. An example is provided by direct instantons \cite{NSVZ}, with 
sizes 
of order $m_Q^{-1}$. 
These contributions  fall off as high powers of $\Lambda_{\rm 
QCD}/m_Q$ (or $\Lambda_{\rm 
QCD}/E$ where $E$ is a characteristic energy release in the
process under consideration), and are very poorly controlled 
theoretically.
Since the fall off of the 
noncondensate nonperturbative corrections is extremely steep, 
basically the  only thing we can say about them is that there is 
a critical value of $m_Q$ (or $E$). For 
lower values of $m_Q$ (or $E$)
no reliable theoretical predictions are possible at 
present. For higher values of $m_Q$  one can  ignore the 
noncondensate nonperturbative contributions. The noncondensate 
nonperturbative contributions are neglected in  the practical  
OPE. In what follows, we will not touch upon these type of effects 
which are associated with the
(small-size) instanton contributions to the coefficient functions. 
There is another, technical, reason why we choose not to 
consider these effects. Since the small-size instantons represent hard 
field fluctuations, all heavy quark expansions carried out in the spirit 
of HQET \cite{HQET} become invalid; the corresponding theory has to 
be developed anew. In particular, the standard HQET decomposition 
of the heavy quark field in the form $Q(x) = \exp\{im_Qv_\mu 
x_\mu\} \tilde{Q}(x)$  becomes inapplicable, as well as the statement 
that all heavy quark spin effects are suppressed by $1/m_Q$, and so 
on. This circumstance is not fully recognized in the literature. 
Due to these reasons, we instead focus on effects due to  
large-size instantons.  This will provide a workable framework
for visualizing the {\em exponential} term. 

At very large $m_Q$ (or $E$), the exponential terms are 
parametrically smaller (in the Euclidean domain) than the power-like 
{\it non}-condensate nonperturbative corrections in the coefficient 
functions. One can argue, however, that this natural hierarchy sets in 
at such large values of momentum transfer where both effects 
are practically unimportant. At intermediate values of the 
momentum transfers -- most interesting from the point of view of 
applications --  an inverse hierarchy  may take place, where the 
exponential terms are numerically more important. 

Ignoring the nonperturbative contributions in the coefficient 
functions is {\em not the only} simplification in the practical OPE. 
The 
series of operators appearing in ${\cal L}$ (the condensate series) is 
infinite.
Practically we truncate it in some finite order, so that the sum in the 
expansion we deal with approximates the exact result, but by no 
means coincides with it. The truncation of the expansion is a key 
point. The condensate expansion is asymptotic \cite{Shif}. Therefore, 
expanding it to higher orders indefinitely
does not mean that the accuracy of the approximation 
to the exact result becomes better. On the contrary, as in any 
asymptotic series, there exists an optimal order. Truncating the 
series at this order, we get the best accuracy. The difference between 
the 
exact result and the series truncated at the optimal order is 
exponential. Large-size instantons, treated in an appropriate way 
will, in a sense,  represent the high-order tail omitted in the 
truncated 
series. 

The essence of the phenomenon -- occurrence of the exponential 
terms -- is similar to the emergence of the condensates at a previous 
stage. Indeed, let us consider, first, the standard Feynman 
perturbation theory.  
At any finite order the perturbative contribution 
is well-defined. At the same time, 
the coefficients of  the $\alpha_s$  series  grow factorially with $n$, 
and 
this means that the $\alpha_s$  series must be, somehow, cut off, i.e.
regularized. The proper way of handling this factorial divergence is
by introducing the normalization point $\mu$ and the 
condensate corrections which 
tempers the factorial divergence of the Feynman perturbative series
in high orders and, simultaneously, 
bring in terms
of order
$\exp {(-C/\alpha_s (m_Q))} $
where $C$ is some positive constant. Loosely speaking, one may say 
that contributions of this
type are  related to the high-order tails of the $\alpha_s$ series.
Similarly, the high-order tails of the condensate (power) series 
correspond to 
the occurrence of the exponential terms. Correspondingly, the OPE, 
even optimally truncated, approximates the 
exact result up to exponential terms. 

The exponential terms not seen in the practical OPE appear both in 
Euclidean, and Minkowski quantities. Their particular 
roles and behaviors are quite different, however. Technically, the 
rate 
of fall off is much faster in the Euclidean domain than  in the 
Minkowski domain, as we will see later. Conceptually, the 
exponential terms in the Minkowski domain determine deviations 
from duality.

Let us now describe what we mean by duality in somewhat 
more detail.
Assume that the effective Lagrangian we work with includes 
external sources, so that the expectation value of this Lagrangian 
actually yields the complete set of physical amplitudes. 
The physically observable Minkowski quantities (i.e. spectra, total 
hadronic widths and so on) are given by the imaginary parts of 
certain terms in the effective Lagrangian. These terms are calculated 
as an expansion in the Euclidean domain. This is  a practical necessity 
-- 
since our theoretical tools are based on the expansions phrased in 
terms of 
quarks and gluons, we have to operate in 
the Euclidean domain.  We then analytically continue in relevant 
momentum transfers to the Minkowski domain. Of course, if we 
could find the {\it exact} result in the Euclidean domain, its analytic 
continuation to the Minkowski domain would yield the exact spectra, 
etc., -- there would be no need in introducing duality at all. In 
reality, 
the calculation is done using the practical OPE. Both, the perturbative 
series 
in the coefficient functions and the condensate series are truncated 
at a certain order. We then analytically continue 
each individual term in the expansion thus obtained, term by term,  
from the deep Euclidean domain to the Minkowski one, and take the 
imaginary part. The corresponding prediction, which can be 
interpreted in terms of quarks and gluons, is declared to be {\em 
dual} to physically measurable quantities in terms of hadrons
provided that the energy release is large. In 
this context dual means approximately equal. The discrepancy 
between the exact (hadronic) result and the quark-gluon prediction 
based on the {\em practical} OPE is referred to as a deviation from  
(local) duality.

Reiterating, in defining duality, we first do a straightforward
analytical continuation, term by term. Defining the analytic 
continuation to
the Minkowski domain in this
way, it is not difficult to see, diagrammatically,
that those lines which were far
off shell in the Euclidean calculation remain hard in the sense that
now they are either still far off shell, or on shell, but carry large
components of the four-momenta, scaling like $m_Q$, or  large
energy release. The sum of the imaginary 
parts obtained in this way is the so called the {\em 
quark-gluon} 
cross section. This quantity serves as a reference quantity in 
formulating the duality relations. When one says that the hadron 
cross section is dual to the quark-gluon one, the latter must be 
calculated by virtue of the procedure described above.

The contributions left aside in the above procedure are related, at 
least at a conceptual level, to the high-order tail of the power 
(condensate) series.  
They can be visualized with large-size instantons. A subtle 
point is that the large-size instantons contribute not only to the 
exponential terms, but also to the condensate (power) expansion.
Our task is to single out the exponential contribution, since we have 
no  intention to use the instanton model to imitate 
the low-order terms of the power expansion. Our instanton model is 
far too crude for 
that. In the next section we proceed to formulating the instanton 
model, 
making special emphasis on this particular element -- isolating 
exponential contributions. 

Beyond the simplest one-variable problems, like the correlation 
function of two vector
currents related to $R_{\rm e^+e^-}$ or $R_\tau$, very often one 
encounters a more complicated situation when the amplitudes have 
several separated kinematical cuts
associated with physically different channels of the given amplitude, 
and one is interested only in one specific
channel. This situation is typical for the inclusive heavy quark 
decays
\cite{CGG,optical}. 
The OPE-based predictions in this case require -- additionally -- a 
different type of 
duality: one needs to assume that a particular cut of interest in the 
hadronic 
amplitude is in one-to-one correspondence with the given 
quark-gluon cut.  In other words, it is assumed that different 
channels (in terms of the hadronic processes and in terms of the 
quark-gluon processes) do not contaminate each other \cite{optical}.  
This was called ``global duality" \footnote{ 
Warning:   the term global duality is 
often used in the literature in a different context.}.
In the practical OPE the cuts of the perturbative coefficient 
functions carry clear identity, and the above assumption of  
``global duality" is easily implementable. 
The above assumption
can actually be proven in the framework of the practical OPE at any 
finite order, 
as was shown in \cite{optical}. However, most probably, this ``global 
duality" 
fails at the level of the exponential terms.  In the present paper we 
do not 
address the issue of  ``global duality" violations
although instantons can model this phenomenon as
well. Such effects are probably smaller than the deviations from local 
duality; in any case they deserve a dedicated analysis. 
It is worth noting that for  one-variable 
problems (e.g. the total $\rm e^+e^-$ annihilation cross section), duality
for various integrals over the cross section over a {\em finite} energy 
range is still local duality, as will be discussed in Sect.~5.

\section{Abstracting General Aspects}

Before submerging into details of the instanton calculations, we 
outline
the practical motivation for inclusion of the corresponding effects 
from 
the general perspective of the short distance expansion. It will also 
enable us to illustrate in a simple way the divergence of the power 
expansion. 
Consider  a generic two-point
correlation function $\Pi(Q^2)$, say, the polarization 
operator for vector currents:
\beq                   
\Pi(Q)\,=\, \int\,d^4x\: {\rm e}^{iqx} \Pi(x)\,=\; 
-\;\int\,d^4x\: {\rm e}^{iqx} \;\langle\: G(x,0)\,G(0,x)\:\rangle _0
\label{a1}
\eeq
where $G$ are the quark Green functions in an external gauge 
field and averaging over the field configurations is implied; we do 
not  
explicitly show the Lorentz indices. Equation (\ref{a1}), and all 
considerations in 
this section, refer to Euclidean space.

The power expansion of $\Pi (Q)$ in $1/Q$ is  the expansion of the 
correlation function $\Pi (x)$ in singularities near the origin. (This 
statement is not quite accurate in the Wilson's OPE; it is correct, 
however, in the practical OPE). Thus, one is interested in the 
small-$x$
behavior of $\Pi(x)$ or, equivalently, of $G(x)$. In the leading, 
deep Euclidean approximation, 
Green's functions are the free ones, $ x\gamma /x^4$, plus 
perturbative
corrections arranged in powers of $\as(1/x^2)$:
\beq
G^{\rm pt}(x)\;=\; \frac{1}{2\pi^2}\,\frac{x\gamma}{x^4}\:
\left(1\,+\,a_1\as(1/x^2)\,+\,... \;\right)\;\;\;\;\;\;\;\; ; \;\;\;
x\gamma = x^{\mu}\gamma_{\mu}
\label{a2}
\eeq
(a particular invariant gauge is assumed here; the correlator is
gauge-invariant anyway, and  similar series can be written directly 
for the product of the two Green  functions). All terms in the 
perturbative 
expansion 
(\ref{a2})
have the same power of $x$, and differ only by powers of 
$\log{x^2}$. 
Logarithms emerge due to the singularity $1/x^2$ of the gluon 
interaction
near $x^2=0$. Upon making the Fourier 
transform,
the perturbative corrections in Eq. (\ref{a2}) are converted into  
powers of $\log{Q^2}$.

Power corrections, $1/Q^n$, emerge from  the expansion of  Green's 
functions near $x=0$: for example, in the Fock--Schwinger 
gauge 
\beq
G(x)\;=\; \frac{1}{2\pi^2}\,\frac{x\gamma}{x^4}\,+\,
\frac{1}{8\pi^2}\frac{x_\al}{x^2}\tilde G_{\al\be}(0) \ga_\be \ga_5 
\;+\;...
\label{a2a}
\eeq
where higher order terms contain higher powers of the gluon field, 
$G_{\alpha\beta}$, or 
its
derivatives at $x=0$ (for a review see \cite{FSch}). Since the
additional terms in the expansion contain 
extra powers of $x$ (generically
 accompanied by $\log{x^2}$), it is clear that, 
returning to the momentum
representation, one gets additional powers of $Q$ in the denominator 
for extra powers of  $x$ in the
expansion of $\Pi (x)$. (The positive powers of  $x$ in  $ \Pi (x)$  are 
accompanied by $\log{x^2}$.) Thus, the $1/Q$ expansion obtained in 
the practical OPE  is in one-to-one correspondence, in the coordinate 
space, with 
the expansion of 
Green's function near the point $x=0$.
(This fact is absolutely explicit in ordinary quantum mechanics, 
where the dynamics are described by a potential. 
In QCD the power corrections to the
inclusive heavy quark decay rates, for example, 
have a similar interpretation: the leading $1/m_Q$ correction,  
due to the Coulomb potential at the position of the 
heavy quark, is absent because of a cancellation
between the initial binding energy and the similar charge 
interaction with the decay products in the final state. 
Physically, the reason is
conservation of color
flow. Moreover, the chromomagnetic term is determined by the 
magnetic interaction at the
origin, and so on.)

The question that naturally comes to one's mind is whether the 
above expansion in the $x$ space is, in a sense, convergent.  At best, 
it can 
have a finite  radius of convergence which, for the
given external field, is determined by the distance to the closest 
(apart from the origin) 
singularity in the complex $x^2$ plane. On the other hand, evaluating
the Fourier 
transform (\ref{a1}) converting $\Pi(x)$ into 
$\Pi(Q)$, one   performs the 
integration over all $x$. Therefore, even for
arbitrarily large $Q$, one has to integrate $\Pi(x)$ in the region 
where 
the  
expansion of the  Green's functions is  divergent. Although any 
particular
power term $1/Q^n$ can be calculated and is finite, this leads to the 
factorial growth
of the coefficients in the $1/Q$ expansion, and thus explains  its 
asymptotic nature.

Let us illustrate this purely mathematical fact in a simplified setting. 
Let us consider the ``OPE
expansion'' of a modified Fourier transform (the 
{\it one}-dimensional 
integral runs from zero to infinity; such transforms are relevant in 
heavy quark theory, see \cite{Shif}):
$$
f(Q)\,=\,\int_0^\infty\,dx\: \frac{1}{x^2+\rho^2}\,{\rm e}^{iQx} 
\;=
$$
\beq
=\;\frac{\pi}{2\rho}e^{-Q\rho} +
\frac{1}{2\rho}[e^{-Q\rho} \overline{\rm Ei}(Q\rho )
-e^{Q\rho} {\rm Ei}(-Q\rho )] \, ,
\label{a3}
\eeq
where $\rm Ei$ is the exponential-integral function.
The integrand has a singularity in the complex plane, at $x = \pm 
i\rho$ and is perfectly expandable at $x=0$. 
Expanding the ``propagator'', $1/(x^2+\rho^2)$, in $x^2$ we
get  the ``OPE series"
\beq
f(Q)\,=\,\int_0^\infty
\,dx\: \sum_{k=0}^\infty\,(-1)^k \frac{x^{2k}}{\rho^{2k+2}}
\,{\rm e}^{iQx}
\;=\;\frac{i}{\rho}\,\sum_{k=0}^\infty\,\frac{(2k)!}{(Q\rho)^{2k+1}}\,
  .
\label{a4}
\eeq
First of all note, that the ``OPE series" has only odd powers of $1/Q$. 
Comparing with the exact expression (\ref{a3})  we see that the  
function $f(Q)$ is 
not
fully represented by its expansion (\ref{a4}), which is obviously 
asymptotic.  The exponential term is missing. This exponential term 
comes from the finite-distance singularities of the integrand.  Indeed, 
one can deform the
contour of integration over $x$ into the complex plane; the integral 
remains the same  as long as the integration contour 
does not wind around the singularity at $x=\pm i\rho$, whose 
contribution 
is
\beq
\de f(Q)\,= \frac{\pi}{\rho}{\rm e}^{-Q\rho} \, .
\label{a5}
\eeq
This is precisely the uncertainty in defining the value of the 
asymptotic 
$1/Q$ series
(\ref{a4}).

Since $f(Q)$ is expanded only in odd powers of $1/Q$, the symmetric 
combination
$g(Q)=f(Q)+f(-Q)$ has no power expansion at all. The function $g(Q)$ 
does not vanish, however:
\beq
g(Q)\,=\, f(Q)+f(-Q)\,=\,\int_{-\infty}^\infty\,dx\: 
\frac{1}{x^2+\rho^2}
\,{\rm e}^{iQx}\;= \;\pi\,\frac{{\rm e}^{-Q\rho}}{\rho}\; .
\label{a6}
\eeq
This expression is in agreement with the above estimate of the 
uncertainty of 
the power
expansion {\em per se} and demonstrates that the exponential terms 
are present.

The appearance of the terms exponential in $Q$ in the example 
above bears some resemblance to the  renormalon issue -- the 
factorial 
growth of the 
coefficients in the
perturbative  expansions in QCD \cite{renorm}. The Feynman graphs 
contain 
integration over all gluon
momenta $k^2$;
on the other hand, the expansion of $\as(k^2)$ in terms of $\as(Q^2)$ 
is {\em
convergent} only for $k^2$ between some minimal and maximal 
scales, $\Lam^2
\lsim k^2 \lsim Q^4/\Lam^2$. Although each particular term in the 
expansion can
be integrated from $k^2=0$ to $\infty$,  yielding a finite number, the 
problem of
divergence of the original expansion of $\as(k^2)$ is resurrected as 
the
factorial growth of the resulting coefficients.

We conclude, therefore, that the divergence of the power expansion 
within the practical OPE, and the presence of the exponential terms 
is a rather 
general phenomenon, and is related, conceptually, to the 
singularities of 
Green's functions in the coordinate space at complex Euclidean values 
of 
$x^2$ located at finite distances from the origin. The question of the 
possible role of these finite 
distance singularities was first raised in Refs. \cite{DS1,DS2}
(see also \cite{Bali}). 

The analogy with renormalons in the $\alpha_s$ perturbative 
expansions can be continued. In the renormalon problem, 
the proper inclusion of the condensates,  within the framework of 
the OPE,  
eliminates the infrared  renormalons altogether
and makes the infrared-related perturbative series well defined and, 
presumably, convergent. One may hope
that a consistent explicit account for the finite-$x^2$ singularities 
would also make the infinite power series well defined. From the 
theoretical perspective, though, this problem has not been 
investigated so far.

Addressing practical applications, there are two general reasons 
to expect that the inclusion of the exponential terms in the analysis 
can be 
important -- even 
though the power series analysis  accounts, at best, for only a few 
leading   
terms in the $1/Q$
expansion. First, there exists some phenomenological evidence, to be 
discussed below, indicating  that the impact of the exponential terms 
in the Minkowski 
domain may be more important numerically than that of the 
omitted condensate terms for intermediate values of the momentum 
transfers.  This statement is illustrated
by the $\tau$ example, see Sect. 5. 
Second, historically, this is not the first case where we 
have  encountered such a perverted hierarchy  in QCD. It is quite 
typical that in the QCD 
sum rules, the contribution of the (omitted) higher order 
$\alpha_s$ terms, which formally dominate over the condensates, is 
far less significant numerically than that of the condensates.  
This point is crucial --  while the exponential terms die out  fast in 
the Euclidean domain, they 
decrease much more slowly in the physical cross sections and, thus, 
often  
dominate 
over the condensate effects, the more so that the latter often are 
concentrated at the end points, and are not seen at all outside the 
end point region.  This observation was emphasized in Ref. 
\cite{DS2}. Indeed, the terms $\sim {\rm e}\,^{-Q\rho}$ 
oscillate, rather than decrease, when analytically 
continued from the Euclidean to the Minkowski domain, $Q \ra i 
E$.

Leaving aside such subtle theoretical questions as the summation of 
the {\em infinite} condensate series, one may hope that including 
the 
principal
singularities at the origin and at finite $x^2$ will lead to  a 
description of the correlation functions at hand which is good 
numerically. Indeed, the leading singularity at 
$x^2=0$
is given by the perturbative  expansion, and subleading terms near 
$x^2=0$ are given 
by the practical OPE.  Adding the dominant singularity at $x^2 \sim 
\Lam^{-2}$ in
the complex plane we capture enough information to describe  the 
main properties of the function,
and thus provide a suitable approximation to the exact result, which 
may work well enough for a wide range of $x^2$, thus yielding the 
proper 
behavior of the
correlation function, $\Pi (Q)$, down to low enough $Q^2$. One should 
clearly realize, 
however, that
this procedure is justified  only if we do not raise  the subtle 
question formulated above and keep just a few of the first 
terms in
both the perturbative  and condensate  expansions. Summing more 
and more condensate terms
within this -- rather  eclectic -- procedure may not only stop 
improving the 
accuracy, but even lead
to double-counting of certain field configurations.
Being fully aware of all deficiencies of this approach at present, we 
still accept it  for  estimating 
possible violations of local duality in a few cases of practical interest. 
Eventually, this approach may develop into a systematic, and 
self-consistent phenomenology of the exponential terms, much in the 
same way as the QCD sum rules represent a systematic 
phenomenology of the condensate terms. 

Technically,  Green's functions are obtained by 
solving the equations of motion in the given background gauge field; 
in 
particular, quark Green's functions are obtained by solving the Dirac 
equation. Henceforth, the singularities outside the origin emerge only 
at  
such (complex) values of $x^2=0$
where the gauge field is singular. Instantons provide an explicit 
example of such fields which are, of course, regular at real $x^2$, but 
have a 
singularity in
the complex $x^2$ plane~\footnote{The gauge potential itself may 
have singularities at real $x^2$, but these are purely gauge artefacts.}. 
The 
singularities of the instanton fields are passed to the Green's 
functions, e.g. the singular terms in the spin-$0$ and spin-$1/2$ 
Green's function have the structure
$$
G_s(x,y) \sim \frac{1}{\left((x-z)^2+\rho^2\right)^{1/2}}\;\;,\;\;\;
\frac{1}{\left((y-z)^2+\rho^2\right)^{1/2}}\, ,
$$
\beq
G_f(x,y) \sim \frac{1}{\left((x-z)^2+\rho^2\right)^\ell}\;\;,\;\;\;
\frac{1}{\left( (y-z)^2+\rho^2\right)^\ell}\; , \;\;     \ell    =\frac{1}{2}, 
\frac{3}{2} \,  ,
\label{a8}
\eeq
where $z$ is the 
center of
the instanton, and $\rho$ is its radius. The actual nature of the 
singularity changes  when one integrates over the position of the 
instanton, and  over its orientation and size.  The poles in $\Pi (x)$ 
may  change into  more complicated singularities (say, cuts). 
Interactions of different instantons will also affect the nature of the 
singularities. One does not expect, however, the finite-distance 
singularities 
to disappear. The precise position of the singularities, and their 
nature,  
depend on the details of the strong dynamics.

The simplest finite $x$ singularity in the physical correlator $\Pi(x-
y)$ one can think of has the form
\beq
\Pi(x)\; = \;\frac{1}{\left(x^2+\rho^2\right)^\nu}
\label{a9}
\eeq
where $\nu$ is some index. (The cases of $\nu =1$ and $2$ were 
discussed in Ref. \cite{DS2}.)
Its momentum representation,
\beq
\Pi(Q^2)\,=\, \int\,d^4x\: {\rm e}^{iQx} \Pi(x)\,=\,
\frac{2\pi^2}{\Gamma(\nu)} \left(\frac{Q\rho}{2} \right)^{\nu-2}
\frac{K_{2-\nu}(Q\rho)}{\rho^{2\nu-4} }
\label{a10}
\eeq
clearly exhibits  the
exponential behavior related to the singularity at $x^2=-\rho^2$. 
The function on the right-hand side is 
exponentially small in the Euclidean domain but yields only an 
oscillating factor (damped by a modest power of $1/Q$)  upon 
analytic continuation to the physical domain, $Q^2=-s -i0$,   
$$
\Im \Pi(s) \;= \;
\frac{\pi^3}{\Gamma(\nu)}
\left(\frac{\sqrt{s}\rho}{2} \right)^{\nu-2}
\frac{\cos{\pi\nu}\,J_{2-\nu}(\sqrt{s}\rho) +
\sin{\pi\nu}\,N_{2-\nu}(\sqrt{s}\rho)}{\rho^{2\nu-4} } \;=
$$
\beq
=\;\frac{\pi^3}{\Gamma(\nu)}
\left(\frac{\sqrt{s}\rho}{2} \right)^{\nu-2}
\frac{J_{\nu-2}(\sqrt{s}\rho)}{\rho^{2\nu-4} }\, .
\label{besselN}
\eeq
Note that at $\nu =1$, the right-hand side of Eq.~(\ref{besselN}) 
implicitly contains $\de(s)$. 
Here, Bessel, McDonald, and Neumann functions are denoted
by $J$,  $K$, and $N$, respectively.
Asymptotically,  at large $s$, 
\beq
\Im \Pi(s)\;\simeq\; 
-\frac{4\sqrt{2}\pi^{5/2}\rho^{3/2}}{s^{5/4}} \frac{1}{\Gamma(\nu)}
\left(\frac{\sqrt{s} \rho}{2}\right)^\nu\; 
\frac{\cos{\left(\sqrt{s}\rho-(\nu+1/2)\frac{\pi}{2} \right)}}
{\rho^{2\nu} }\;\; .
\label{a11}
\eeq
Certainly, the purely oscillating factor above is an extreme case. 
Any sensible smearing over $\rho$ (with a smooth weight function)
will restore the 
decrement of the exponent in the Minkowski domain (as discussed 
in Ref. \cite{Shif}). Generically, therefore, we obtain a decaying 
exponent, $\exp (-E^\sigma )$, modulated by oscillations. More 
exactly, we get a sum of such terms. The index $\sigma$ depends on 
dynamics and, in principle, can be rather small numerically. If 
$\sigma$ is small, the damping regime takes over the oscillating 
regime at large values of $E$, after a few  unsuppressed 
oscillations occur.  At such values of $E$, the powers of $1/E$ in the 
pre-exponent can make the whole contribution small. Therefore, 
starting our analysis with an extreme situation --
a purely oscillating factor times some power of $1/E$ in the pre-
factor 
-- is quite meaningful. We will discuss all these details in a more 
specific setting of the instanton model. 

It is easy to see that causality requires 
singularities of $\Pi (x^2)$  to lie either on the negative real axis of 
$x^2$, or at
larger arguments of $x^2$, on the unphysical sheet; $\Pi(x^2)$ must 
be 
analytic at $|\arg{x^2}|< \pi$.
In the instanton model the singularities are on the 
negative  $x^2$ axis. Smearing  the instanton sizes with a
smooth function weakens the strength of the singularity near the
purely imaginary $\sqrt{x^2}$ and thus effectively moves it further
into the complex plane.
                      
To summarize, we argued  that the violations of local duality are
conceptually related to the divergence  of  the condensate 
expansion (practical OPE) in high orders. Technically,  they may
occur due to  the  singularities of  Green's functions at complex
Euclidean values of $x$  at finite distances from the origin.
Accounting for such singularities, in addition to the perturbative 
and the condensate expansion, is a  natural first step beyond the
framework of the practical OPE. In the next section we will proceed to a
specific model for this  phenomenon based on instantons. Since they
are not necessarily the  dominant vacuum component we try to limit
our reliance on  instantons to the absolute minimum.  In particular,
their topological  properties are inessential for us, and even lead
to certain superfluous  complications.

\section{Instanton Model}

Here we will formulate our rules of the game.
To get an idea of possible violations of duality we will consider a set 
of physically interesting processes (two-point functions of various 
currents built from light quarks, the transition operators relevant 
for the inclusive heavy quark decays and so on). Our primary goal is 
isolating the finite-distance singularities in $x^2$, which will  
eventually be converted into the exponential terms in the 
momentum 
plane. To this end it will be assumed that the quark Green's functions 
in 
the amplitudes under consideration are Green's functions in the 
given one-instanton background. The one-instanton field is selected 
to represent coherent gluon field fluctuations for technical reasons 
-- in this background Green's functions for the massless quarks are 
exactly known. 

The instanton field depends on the collective coordinates -- its 
center, color space orientations, and its radius. Integration over all 
coordinates except the radius is trivial, and will be done 
automatically.  Integration over the instanton radius $\rho$ requires 
additional comments. 

First of all, in all expressions given below, integration over  $\rho$ is 
not indicated explicitly unless stated otherwise. Any expression 
$F(\rho )$ should be actually understood as follows
\beq 
F(\rho ) \ra \int \frac{d\rho}{\rho} d(\rho ) F(\rho ) 
\eeq
where $d(\rho )$ is a weight function and integration over instanton 
position,
$d^4z/\rho^4$, is included in the definition of $F(\rho ) $. 

If we were building a dynamical model of the QCD vacuum based on 
instantons, we could have tried to calculate this weight function. As a 
matter of fact,
for an isolated instanton the instanton density, $d(\rho )$, 
was found in the 
pioneering work \cite{t'Hooft}; for pure gluodynamics, 
\bb
 d(\rho)_0 = {\rm const}\, 
 (\rho\Lambda_{\rm QCD}) ^b,
\ee  
where $b$
is the first coefficient in the 
Gell-Mann-Low function ($b = 11/3$ $N_c$ for the $SU(3)$
gauge group). Of course, the approximation of the instanton gas
\cite{gross} is totally inadequate for many reasons -- one of them is 
the fact that inclusion of the massless quarks completely suppresses
the isolated instantons \cite{t'Hooft}. This particular drawback can be 
eliminated if one takes into account the quark condensate, $\langle 
\bar q q\rangle \neq 0$. Then the instanton density takes the form
\cite{shifvainst1}
\bb
d(\rho)\; =\; {\rm const}\: (\langle \bar q q\rangle \rho^3)^{n_f} d_0 
(\rho)
\ee  
where $n_f$ is the number of the massless  quarks, and now $b = 
11/3 
N_c - 2/3 n_f$. Note the extremely steep $\rho$ dependence  of 
the instanton density at small $\rho$. The impact of the quark 
condensates is not the end of the story, however, since for physically 
interesting values of $\rho$, the vacuum field fluctuations form a 
rather dense medium
where each instanton feels the presence of all other fluctuations. 
In principle, one could try to build a model of the QCD vacuum in this 
way, for instance, that is what is done in the
 so called instanton liquid model (see
\cite{liquid,Dodik} and references therein). The main  idea is that the 
instanton density is sharply peaked 
at  $\rho\approx 1.6 \,\, ({\rm GeV})^{-1}$, where 
the classical action is still large, 
i.e. we  can still  consider  individual instantons. 
On the other hand, the interaction
between  instantons is also large, but still not large enough so that
the 
instantons
 melt. The extremely steep growth of the instanton density at small 
$\rho$ is cut off abruptly at larger $\rho$, due to  interactions in the 
instanton liquid. 
The proposed model density which captures these features is just  
 a plateau at $\rho=\rho_c$ with the  width 
$\delta\ll\rho_c$.

We would like to avoid addressing dynamical issues of the QCD 
vacuum in the present paper. Our task is to rely on general features, 
rather than on  specific details, and the instanton field,  for us, is 
merely representative of a strong coherent field fluctuation. For 
this limited purpose, we can ignore the problems of the calculation 
of the instanton density, and just postulate the weight function
$d(\rho )$ in the  simplest form possible.
The most extreme assumption is to approximate $d(\rho )$
by a delta function,
\beq
d(\rho ) = d_0\;\rho_0 \delta (\rho - \rho _0 ),
\label{deldens}
\eeq
where $d_0$ and $\rho _0$ are  appropriately chosen constants.
In a  very crude  approximation this weight function is 
suitable, in principle, although it has an obvious drawback. 
If $\rho$ is fixed,
as in Eq. (\ref{deldens}), 
the instanton exponential in the Euclidean domain becomes cosine in 
the Minkowski domain, with no decrement. 
For instance, $(Q\rho )^{-1}K_1 (Q\rho )\ra  E^{-3/2} \cos{(E\rho - 
{\rm phase})}$, where the arrow denotes continuing to the 
Minkowski 
domain, taking the imaginary part, and
keeping the leading term in the expansion for large $ E\rho$.
 If one wants to be more realistic, one 
should introduce a finite width. A reasonable choice might be
\begin{equation}
w(\rho ) = (\rho )^{-1} {\cal N} \exp\{-\frac{\alpha}{\rho}-
\beta\rho\}
 \label{fsid}
\end{equation}
where ${\cal N}$ is a normalization constant,
\begin{equation}
\alpha=\frac{\rho_0^3}{\Delta^2},\,\,\,\, \beta=  
\frac{\rho_0}{\Delta^2} ,
\label{constants}
\end{equation}
$\rho_0$ is the center of the distribution, and $\Delta$ is its width.
Convoluting $(Q\rho )^{-1}K_1 (Q\rho)$ with this weight function,
one smears the cosine, which results in the exponential fall off
in the Minkowski domain,
\begin{equation}
(Q\rho )^{-1}K_1 (Q\rho )\ra
{\cal N} 2E^{-1}J_1
\{ \sqrt{2\alpha}[\sqrt{\beta^2+E^2}-\beta]^{1/2}\}
K_1\{\sqrt{2\alpha}[\sqrt{\beta^2+E^2}+\beta]^{1/2}\}
\label{usrosc}
\end{equation}
where the meaning of the arrow is the same as above. 

If
\begin{equation}
E\gg \frac{1}{\Delta}\frac{\rho_0}{\Delta},
\label{egg}
\end{equation}
the  imaginary part reduces to
\begin{equation}
2{\cal N} E^{-1} J_1 (\sqrt{2E\rho_0}\frac{\rho_0}{\Delta} )K_1
(\sqrt{2E\rho_0}\frac{\rho_0}{\Delta} )\, ,
\label{asus}
\end{equation}
and falls off exponentially. If the weight function is 
narrow, ($\Delta\ll\rho_0$), this exponential suppression starts at 
high energies,
see Eq. (\ref{egg}). In the limit when $\Delta\rightarrow 0$, with $E$ 
fixed, the exponential suppression disappears from Eq. (\ref{usrosc}), 
and we return  to the original oscillating imaginary part. Note 
also that the exponent at $E\gg \frac{1}{\Delta}\frac{\rho_0}{\Delta}
$ is different
from the one in the Euclidean domain ($\sqrt{E}$ {\em 
versus} $Q$).  In Sect. 5.2 we will introduce the corresponding index, 
$\sigma$, characterizing the degree of the exponential fall off in the 
Minkowski domain at asymptotically large energies. 

Concluding this section, we pause here to make two  remarks of 
general character. The fact that smearing
the scale with smooth functions of the type (\ref{fsid})
produces exponential fall off 
is not specific to the instanton-induced spectral density. Even much 
rougher spectral densities (with appropriate properties), being 
smeared with the weight function (\ref{fsid}), become exponential. 
An instructive example is provided by a model spectral density
suggested in Ref. \cite{Shif}. Consider the following ``polarization 
operator"
$$
``\Pi " \propto \beta
\left(\frac{Q^2 +\Lambda^2}{2\Lambda^2}\right)
$$
where $\beta$ is the special beta function  related to Euler's $\psi$ 
function,
\beq
\beta (x) =\frac{1}{2}\left[ \psi\left(\frac{x+1}{2}\right)
- \psi\left(\frac{x}{2}\right)\right] =\sum_{k=0}^\infty\frac{(-
1)^k}{x+k}\, . 
\label{fakepo}
\eeq
This fake polarization operator mimics, in very gross features, say, 
the difference between the vector-vector and axial-axial two-point 
functions \footnote{In Ref. \cite{Shif} the model spectral density 
(\ref{fakepo})
was suggested in the context  of the heavy-light two-point functions, 
and, accordingly,  $x$ was related to $E$, not $Q^2$.  
In the heavy-light systems, the model does not reproduce fine 
features either; in particular, the
equidistant spectrum it yields is not realistic. The separation 
between 
the highly excited states should fall off as $1/E$. Such a behavior 
immediately follows
from the semiclassical quantization condition 
$$
\int{(E-\Lambda^2 r)} dr \propto n\;\;.
$$ 
Previously this pattern  was noted in the 
two-dimensional 't Hooft model \cite{Zhit}.}.  At positive $Q^2$, 
it is expandable in an 
asymptotic series in $1/Q^2$, plus exponential terms. At negative 
$Q^2$, (positive $s$), it develops an imaginary part. 
The imaginary part  obviously consists of two infinite combs of 
equidistant delta functions -- half of them enter with the coefficient 
1, the other half with the coefficient $-1$. Literally speaking, there is
no local (point-by-point) duality at any energies. 

Let us smear the combs of the delta functions with the
weight function (\ref{fsid}). Now, the imaginary part at negative $x$
is smooth, exponentially suppressed, and oscillating, 
\beq
{\rm Im}\, \int_0^\infty d\rho \; w(\rho)\, \beta (\rho x )\; =\;
\frac{\pi}{x}\sum_{k=1}^{\infty}\: (-1)^k 
w\left(k/|x|\right)\;
 \propto \;
\Im \, {\rm e}\,^{-\sqrt{\pi/2}\,(1-i)\,\sqrt{\alpha|x|} 
\:+{\rm const} }\; . 
\eeq
Indeed, one can represent the sign alternating sum, $(-1)^k
w(k/|x|)$, as the integral of the function $i/(2\sin(\pi z)\,w(z/|x|))$,
with complex variable $z$, over the contour embedding the positive 
real
axis $[1, +\infty)$.  
At large $|x|$, its value is determined by large
$z$; the integrand has two complex conjugated saddle points,
$z=\sqrt{\frac{\alpha|x|}{\pi}}\,{\rm e}^{\pm i\pi/4}$, whose 
steepest
descents lead to $z=0$ and $z=\pm i\infty$. Evaluating the saddle 
point
integrals, one arrives at the above asymptotics.

Returning to the instanton model, we note that the 
weight function, (\ref{fsid}),
is  convenient, because 
the contribution of the small-size instantons (which affect the OPE
coefficients and are not discussed in the present paper)
are naturally suppressed.
The absence of these small-size instantons allows for a sensible  
expansion
parameter, $1/(m_Q\rho)$,   which can be used in calculations 
with heavy quarks.  It is worth emphasizing again that at very 
large momentum transfers (energies, heavy quark masses, etc.),
the small-size instantons will always dominate over the exponential 
terms. Thus, our model is applicable, if at all, only to intermediate 
scales. 

In QCD, the instanton field configuration does not constitute any 
closed approximation. Therefore, one may question practically every 
aspect of the model we suggest. Developing phenomenology of the 
exponential terms will help us understand whether this approach has 
grounds. 
From the purely  theoretical standpoint it might be instructive  to 
consider a formulation of the 
problem where the  instantons can be studied in a clean 
environment, rather than in the complex world of QCD. Such an 
analysis was already outlined in the literature \cite{MS}.
Let us assume that instead of QCD, we study the Higgs phase, i.e.
we introduce scalar colored fields which develop a vacuum 
expectation value, and break color symmetry spontaneously. 
The gluon fields acquire masses. If their masses
 are much larger than
$\Lambda_{\rm QCD}$, we are in the weak coupling regime, and the 
semiclassical approximation becomes fully justified. The instanton
contribution to various amplitudes is well-defined now, and subtle 
questions, which could not be reliably answered in QCD, can be 
addressed.

The pattern of the instanton contribution as a function of energy  
in this case was studied in Ref. \cite{MS}.  It is quite remarkable that 
the pattern obtained bears a close resemblance to what we have in 
QCD, in particular, oscillations in the Minkowski domain. 

 \section{Deviations from duality in $R_{\rm e^+e^-}$ or hadronic $\tau$ 
decays}

\subsection{Instanton estimates}

The peculiar details of local duality violations are more 
transparent in the simple cases of  $\rm e^+e^-$ annihilation 
cross section
and inclusive 
hadronic $\tau$
decays. Several  rather sophisticated analyses of the 
instanton
effects in these problems were carried out recently 
\cite{bal,balbb,nason,?}.  A more 
general
consideration, rather close in ideology to our approach,
 was given   in 
\cite{DS2} (in a sense the spirit of the 
suggestion of Ref. \cite{DS2} is  more 
extreme). We further comment on these works  
in Sect.11.
To see  typical features of the instanton-like effects we consider, 
for
simplicity, the correlator of the flavor-nonsinglet vector currents 
relevant to 
$R_{\rm e^+e^-}$; a similar correlation function appears in the hadronic 
$\tau$
decays, alongside with its axial-vector counterpart. For simplicity, we 
will mainly ignore the latter contribution and discuss 
the vector
part as a concrete example.

Let us define ($Q^2=-q^2-i0$) 
$$
\Pi_{\mu\nu}(q^2)\,=\,  \,\int\,d^4x\: {\rm e}^{iqx}
\;\langle\;iT\left\{J^+_\mu(x)\, J_\nu(0)\right\}\; \rangle\;=\; 
\frac{1}{4\pi^2}\,(q^2\de_{\mu\nu}-q_\mu q_\nu)  \Pi(q^2)\, ,
$$
\beq
\Pi(Q^2)\;=  \;\log{\frac{Q^2}{\mu^2}} \;+\;...\,\, ,
\eeq
$$
R(s)\;=\;-\frac{1}{\pi}\, \Im \Pi(Q^2=-s-i0) = 1\;+\; ... \; .
\label{b1}
$$
For the purpose of our discussion the average $\langle...\rangle$ is 
not yet
understood as averaging over the physical vacuum; we, rather,  
calculate the correlation function in a 
particular
external field, and average over  certain parameters of 
this
field (the invariant tensor decomposition is appropriate in the latter 
case).
The second equation shows $\Pi(Q^2)$ in the absence of any field. In 
a
given field, $\Pi_{\mu\nu}(x,y)$ is merely a trace of the product of 
the two
Green functions which are explicitly known for massless quarks in 
the field of
one instanton. Upon averaging over the positions and orientations of 
the 
instanton of the fixed size $\rho$, one arrives at the known 
expression (integration over $\rho$ is shown explicitly) 
\cite{oneinstR,DS1}
$$
\Pi(Q)\,=\,\Pi_0(Q)\,+\,\Pi^{\rm 
I}(Q)\, =
$$
\beq
 \;= \,\log{\frac{Q^2}{\mu^2}} 
\;+\;  
16 \pi^2 \,\int \frac{d\rho}{\rho} d(\rho) 
\:\left[\frac{1}{3(Q\rho)^4}- \frac{1}{(Q\rho)^2}
\int_0^1\,dt\: K_2\left(\frac{2Q\rho}{\sqrt{1-t^2}}\right) \right]
\label{b2}
\eeq
where $K_2$ is a McDonald function, and $ \int 
(d(\rho)/\rho^5)d\rho$ 
is to be identified with the number of instantons per unit volume. 
The superscript I marks the instanton contribution. 
The first
term in the square brackets is ``a condensate", the second one, on the 
contrary, does
not produce any $1/Q^n$ expansion. Considering \eq{b2} in the 
Minkowski domain
one has ($E=\sqrt{s}$)  
$$
R(E)\,=\,R_0(E)\,+\,R^{\rm I}(E)\,=
$$
\beq
1 \;+\;
8 \pi^2 \int \frac{d\rho}{\rho} 
d(\rho)\:\left[\frac{1}{2\rho^2}\de(E^2)+ \frac{1}{(E\rho)^2}
\int_0^1\,dt\: J_2\left(\frac{2E\rho}{\sqrt{1-t^2}}\right) \right]\;\;,
\label{b3}
\eeq
where $J_2$ is a Bessel function. Violation of local duality at 
finite $E$
is given by the last term. Expanding the Bessel function
at large $E\rho$, and performing the saddle point
evaluation of the inner integral, we see that
it oscillates, but decreases in magnitude 
only as $1/E^3$:
\beq
R^{\rm I }(E)|_{fixed\,\,\rho} \,\simeq\;
-\:4 \pi^2 \,\frac{1}{(E\rho)^3}\cos{(2E\rho)}\, ,
\label{b4}
\eeq
(we remind the reader that the true power corrections from the OPE 
appear in the 
imaginary part at large $E$  only at the 
level $\alpha_s^2/E^4$ provided that the quarks are massless, as
we assume here).
After averaging this result over $\rho$ with a smooth enough 
weight, the resulting $R^{\rm I }$ decreases exponentially at 
$E\ra
\infty$; the decrement is determined by the analytic properties 
of
$d(\rho)$. The behavior of the ``exponential'' contribution, given by 
the last
term in \eq{b3} at small $E$, is relatively smooth,
\beq
R^{\rm I }(E)\;\sim \;-
4 \pi^2 \, \log{(E\rho)}\;\;.
\label{b5}
\eeq

\begin{figure}
  \epsfxsize=12cm
  \centerline{\epsfbox{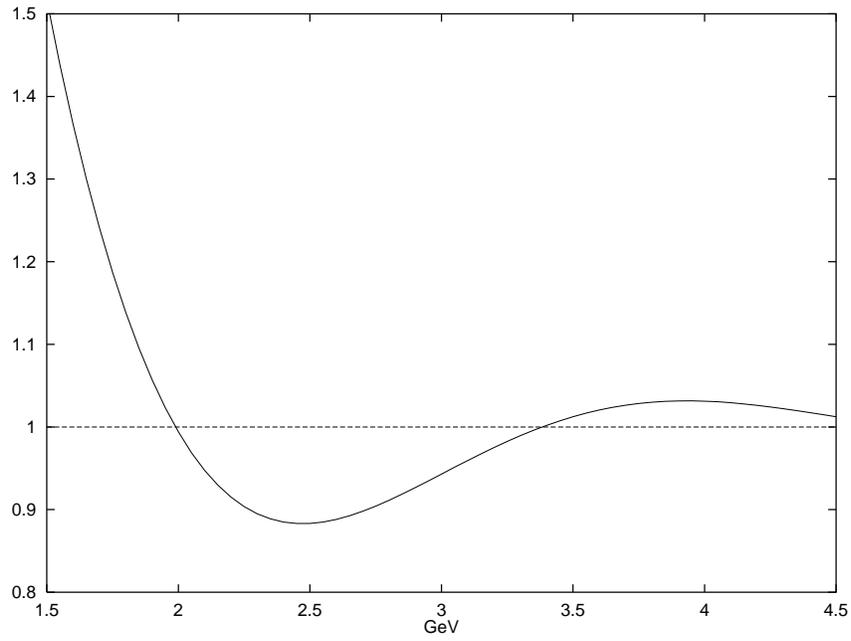}}
  \caption{ $R(E)$, taking into account the instanton contribution.
            The perturbative result is normalized  to unity. }
  \label{fig1}
\end{figure}

To visualize the above 
expressions, we plot in Fig.\ref{fig1}      
 the value of 
$R(E)$ stemming from Eq. (\ref{b3}) 
in the real scale $E$ using the instanton density 
(\ref{deldens}), with some rather {\it ad hoc} overall normalization
\footnote{In Sect. 9.1 we discuss our choice for $d_0$. 
 Our motivation is based on  phenomenological 
analysis of the duality violations in the semileptonic $D$ decays,
which may be as large as $\sim 50$\% .}
$d_0$, and 
 $\rho_0=1.15\GeV^{-1}$.

\begin{figure}
  \epsfxsize=12cm
  \centerline{\epsfbox{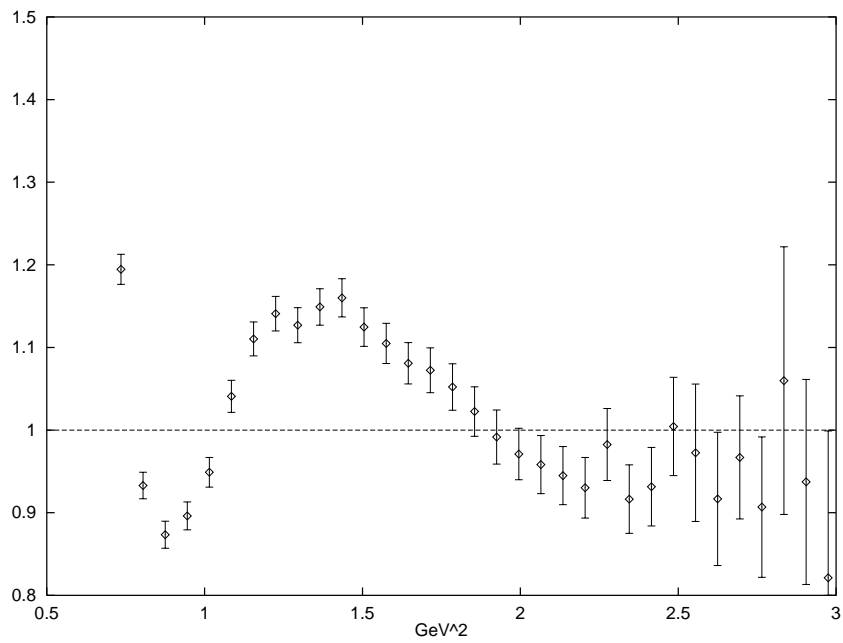}}
  \caption{Experimental value of $R(E)$.  }
  \label{fig2}
\end{figure}    

  Figure \ref{fig2} represents experimental values
extracted from the CLEO data
 \cite{data}. 
Although our theoretical curve does not 
{\em literally} coincide with the actual data, it is definite that the 
general 
feature of
the experimental curve -- the presence of oscillations, and a
moderate falloff of their magnitude with energy -- is captured 
correctly. The fact that our model is not accurate enough to ensure
the point-by-point coincidence  was to be 
anticipated. Obvious deficiencies of the model will be 
discussed in Sect. 10.  Some additional remarks concerning duality 
violations in the hadronic $\tau$ decays are given in Sect. 9.1, see
\eq{tauwid}.

\subsection{Three zones}

A single glance at experimental data (a part of the data
is presented on Fig.  2) reveals a striking regularity
of the inclusive cross section. We believe that this regularity
is a general phenomenon, and its discussion is very pertinent 
to the issue of the duality violations. The same pattern of behavior is 
expected, say, in the spectra of the radiative decays $B\rightarrow 
X_s + \gamma$, and so on.

One can single out three distinct zones in the physical inclusive 
hadronic cross sections, governed by different dynamical regimes.
If we proceed from the low invariant masses of the inclusive 
hadronic state to high masses, the first zone we see is a 
``narrow resonance" zone. It includes one, or at most two, 
conspicuous 
resonances. It stretches up to a first boundary -- call it $s_0$. 
Crossing this first boundary, we find ourselves in the 
 second zone --  the oscillation zone. The cross section here is already 
smooth, and the point-by-point violations of the quark (gluon) 
hadron duality are not violent. Still, these violations are quite
noticeable (they may constitute a few dozen percent),  and have 
a 
very clear pattern --  several
clearly visible oscillations, with relatively mild
suppression,
$$
R= R_{\rm OPE} + ({\rm const}/E^k)\sin (2E\rho + \phi)\;\; .
$$
The upper boundary of this 
zone will be referred to as $s_1$. Finally, above 
this second boundary, 
there lies a third domain -- the asymptotic zone, where 
$$
R= R_{\rm OPE} + \exp [-(2E\rho)^\sigma] \sin{\left((2E\rho')^\sigma
 + \phi\right)}
\;\;\; {\rm or}\;\;\;  (1/E)^\gamma\,, \;\;\; \sigma < 1\;,\;\;\;
\gamma \gg 1\;\; .
$$
Here $R_{OPE}$ is a smooth (practical) OPE prediction, $k$ is an 
integer,
$\sigma$ and $\gamma$ are indices. Our model is intended for 
applications in  the second (oscillation) zone.

It is worth noting that the precise values of the boundaries
$s_0$ and $s_1$ are very sensitive to dynamical details. For instance,
in the imaginary world with infinite number of colors,
$s_0$, is believed to go to infinity, and the regime of the second zone 
never occurs.

\subsection{Smearing and local duality.}

In this section, we discuss another general, and crucial feature of the 
``exponential" terms. What happens if, instead 
of 
considering the imaginary parts point-by-point, we choose
to analyze some integrals over a finite energy interval,  
with some weight? 
Intuitively, it is clear that violations of 
the quark (gluon) -- hadron duality are expected to become 
 smaller if the weight function is smooth enough 
and the
energy interval over which we integrate is large. The case when one 
integrates with a 
polynomial weight (polynomial in 
$s=E^2$) is of a particular practical interest. Let us consider the 
finite-energy moments
\beq
{\cal M}_n(s)\,=\,(n+1)\,\int_0^s\: R^{\rm I }(t)\,t^{n}\,dt\;\;.
\label{b6}
\eeq

The deviations from duality are smallest at the upper edge of the 
integration domain, and largest at the lower edge.  Intuitively,
it is 
clear that the deviations from duality in the integral (\ref{b6})
are determined by deviations at the upper edge of the integration 
domain.
This result, however, is obtained only if one makes full use of the
analytic properties of the exponential contributions at hand.
If one tries to directly integrate the asymptotic instanton formulae
over $t$, in a straightforward manner, one gets a huge contribution 
determined by the lower end. This is the essence of the so called
``a part larger than the whole" paradox, observed in the instanton 
calculations, say, in Refs. \cite{Chy,AFalk}, where
the instanton contribution to the decay spectrum turned out to be 
parametrically larger than that to the total decay rate. 

Let us elucidate the point in more detail. 
The moments, ${\cal M}_n(s)$, get contributions both from the usual 
OPE terms, which are located at small $s \sim \Lam^2$ 
(in our case it is 
$\de'(s)$ from the term $1/Q^4$ in \eq{b2} which 
survives only for $n=1$), and from the
exponential part going beyond the practical OPE. We are interested 
here only
in the latter piece and, therefore,  subtract the  condensate part. 
Using the large-$s$ expansion for $R$, 
one is literally in trouble: the integral over the imaginary part 
(\ref{b4})
seemingly diverges at small $s$ where this expression is not 
applicable, and must
be cut off at  $s_0 \lsim 1/\rho^2$. 
At first sight, it  then seems the
result  completely
depends
 on the lower limit $s_0$, and on the precise way of implementing 
the 
cut off at $s_0$. The fact that we integrate over a large interval 
stretching up to $s$ seems to be of no help in suppressing the 
duality violations.  

It is easy to
see, however, that 
the large result above is obtained only because we have used a 
wrong 
expression at small $s$. The asymptotic instanton formula is 
definitely invalid at small $s$. If $\rho$ is fixed, we could use, of 
course, the exact instanton expression at small $s$ which is 
(almost)
not 
singular (see Eqs. (\ref{b5}) and (\ref{b3})). We would not trust the 
instanton result 
at small $s$ anyway. Therefore, the prediction for the moments,
${\cal M}_n(s)$,
should be obtained without relying on the expicit expressions at 
small $s$. To  this end  one invokes dispersion relations.

 Whatever the origin of the exponential contribution under 
consideration is, it must obey the dispersion relations. 
Take  $\Pi (Q)-\Pi_{\rm OPE} (Q)$, where 
the 
``practical OPE'' piece, $\Pi_{\rm OPE} (Q)$, in our example 
is given explicitly by the single term
\beq
\Pi_{\rm OPE} 
  (Q)\;\equiv\;\frac{16\pi^2 }{3\rho^4}\:\frac{1}{Q^4}\;\;.
\label{b8}
\eeq
(we omit the subscript $\rm I$, since,
 in what follows we consider only 
instanton
induced contributions).
Since $\Pi (Q)-\Pi_{\rm OPE} (Q)$ exponentially 
decreases
at large Euclidean $Q^2$, one has an infinite number of constraints
\beq
\lim_{Q^2\ra\infty}{\;Q^{2n}
\left(\Pi (Q)-\Pi_{\rm OPE} (Q) \right)}\;=\;(-
1)^{n}\;
\int_0^\infty\;\left(R (s)-R_{\rm OPE} (s)\right)
\,s^{n-1}\,ds\;=\;0\;\;.
\label{b9}
\eeq
In other words, all moments considered in the full $s$ range from 
$0$ to
$\infty$, ${\cal M}_n(\infty)$, are given completely by their 
OPE values, and the
extra contribution from the non-dual piece is absent.  (To define
${\cal M}_n \equiv {\cal M}_n(\infty)$ in  the particular example one
may need  to regularize  integrals in (\ref{b6}), (\ref{b9}) by, say,
introducing a damping exponent  ${\rm e}^{-\epsilon \sqrt{s} }$ with
an infinitesimal $\epsilon$.) 

Using this property, one immediately concludes that the violation of 
duality in the moments, ${\cal M}_n(s)$, is determined, 
parametrically, by 
the {\em upper} limit of integration $s\,$:
\beq
{\cal M}_n(s)\;=\;{\cal M}_n^{\rm OPE}(s) \,- \,(n+1)\, 
\int_s^\infty\;R  (t)
\,t^{n}\,dt\;\;.
\label{b10}
\eeq
This is, clearly, the most general property of the ``exponential'' terms 
which
does not depend on any details of a particular ansatz.

Formally, the relations of the type (\ref{b9}) and  (\ref{b10}) for the 
imaginary  part (obtained by the 
 analytic continuation of the Euclidean exponential  terms)  can be 
written  as follows:
\beq
R (s)-R_{\rm OPE}(s)\;=\;
\int_0^\infty\;dt\: \Phi (t) \left[\de(t-s)-
{\rm e}\,^{-t\frac{\partial}{\partial s}} \de(s)\right],
\label{b11}
\eeq
where $\Phi(t)$  vanishes at $t\le 0$ and coincides with the 
asymptotic instanton expression at positive $t$,  
\beq
\Phi(s)\;=\;-4\, \pi^2\,  \; \frac{1}{(\rho \sqrt{s})^3} 
\cos{(2\rho \sqrt{s})}\;+\; {\cal O}\left(s^{-2}\right)\;\;.
\label{b12}
\eeq
In other words, to do the smearing integrals properly one must 
substitute $R(s)$ by $R(s)$ plus the whole tower of 
terms presented on the right-hand side of Eq. (\ref{b11}). 

The representation (\ref{b11}) is convenient since it explicitly 
ensures the property (\ref{b10}), 
which is the fact that the corresponding 
contribution
to the correlator dies out faster than any power in the deep 
Euclidean domain.
It shows that any particular spectral density  generated at 
large $s$ as
a violation of local duality,  must be necessarily accompanied by the
corresponding OPE-looking terms located at small $s$; disconnecting  
these 
seemingly
different contributions  is not consistent with 
analyticity.

Let us parenthetically note that
 similar relations, with delta functions at the end point,  must be 
used in the instanton calculations  of the
semileptonic spectra in the heavy quark decays. The occurrence of 
the 
end-point delta functions in the instanton expressions is reminiscent 
of what happens with the regular (OPE) power 
corrections to the 
semileptonic widths \cite{prl}. The interaction with the final quark
magnetic
moment does contribute to the inclusive lepton spectrum with a 
definite sign 
in its regular part. And, yet, it is known to be absent in the total 
width. 
The 
cancellation occurs due to the terms located at the end point of 
the
spectrum which -- in the naive approach -- are not seen in 
the $1/m_Q$
expansion.

Loosely speaking,  the part referring to low $t$ in the integral 
(\ref{b6}) is eaten up  by the  ``condensates''. 

\Eq{b10} demonstrates that the deviations from  duality in the 
finite-energy
moments, ${\cal M}_n (s)$, are generically given by the accuracy of 
local 
duality at the
maximal energy scale covered, $s$. More exactly, the error is 
approximately given by
the integral over the last half-period of oscillations. 
For the sign-alternating
combination of the moments, similar to the one determining the 
hadronic width
of the $\tau$ lepton, 
$$\Gamma_{\rm had}(\tau)\sim 
2{\cal M}_0(m_\tau^2)/m_\tau^2-
2{\cal M}_2(m_\tau^2)/m_\tau^6+ {\cal M}_3(m_\tau^2)/m_\tau^8\, ,
$$
it is likely to be  larger 
and can be governed by a lower scale. Indeed, 
 the resulting weight function 
\beq
w_\tau(s)\;=\;2\; \vartheta(m_\tau^2-s) \cdot
\left(1+2\frac{s}{m_\tau^2}\right)\left(1-\frac{s}{m_\tau^2}\right)^2
\label{b13}
\eeq
is saturated mainly at $s\lsim m_\tau^2/3$. Therefore, $s\sim 
m_\tau^2/3$ can be viewed as the actual mass scale 
governing the
duality violation in this problem.  Numerically it is close to 
the first pronounced resonance in the axial channel.

An obvious  reservation is in order here. In  real QCD, where the 
series of the power
corrections in the OPE is infinite, and presumably  factorially divergent,  
untangling the exponential terms from the high-order tail of the 
series remains obscure.  There is no answer to the question 
``what is the  
summed infinite OPE series?", even in the
Euclidean domain. Our approach to this issue is purely
operational, 
and is clearly formulated in simple problems:  pick up the
contribution of the finite $x$ gularity in the saddle point
approximation. It is motivated by the 
general consideration of Sect.~3. 

\section{Soft instantons in the $1/m_Q$ expansion.}  

In this section, we briefly outline the generalities of the instanton 
induced
exponential corrections of heavy quark decays. 
The  goal of this section is a ``back of the envelope" 
calculation presenting the functional dependence on the 
heavy quark mass. More detailed 
calculations, which will provide us with all coefficients 
in the pre-exponential factors, are deferred until Sects. 7 and 8.
 
The main feature of the problems at hand 
 is the presence of a large parameter, $m_Q\rho$, 
which allows us to obtain sensible analytic expressions.
As we have already discussed in a general context, there are 
three types of contributions associated with 
instantons:
(i) Small size instantons affect the coefficient functions of the OPE.
    We are not interested in these terms. They will not appear
   in our calculations, since instantons of small size are,
 by definition,
excluded from our model density function $d(\rho)$, and we always 
assume that
$m_Q\rho\gg 1$. Technically, as was already mentioned,  small
size instantons cannot be taken into account using the standard
methods of HQET.
(ii) The terms proportional to powers of $1/(m_Q\rho$). They 
represent  the instanton
contributions  to the matrix elements of various  finite-dimension 
operators that
are present in the OPE. (In the present context these terms are actually
pure 
contamination, and so we will discuss only how to get rid of 
them). (iii) Finally, there are exponential terms, of 
the 
form 
$\exp{(-2m_Q\rho)}$.  These terms are our focus.  

\subsection{Decays into light quarks}

\begin{figure}
  \epsfxsize=12cm
  \centerline{\epsfbox{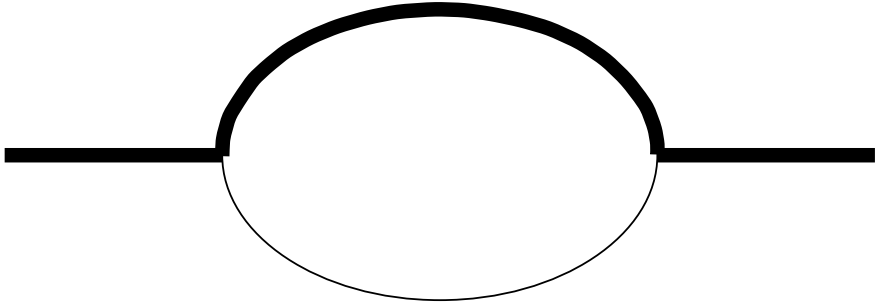}}
  \caption{Forward scattering amplitude. Bold lines represent 
          propagation
          of a  particle in the instanton field. }
  \label{fig3}
\end{figure}

Consider the generic form of an inclusive forward scattering 
amplitude which corresponds to the decay of a heavy quark into
a massless quark and a number of color singlet particles, say, $l\nu$ 
(Fig.\ref{fig3}),

\bb
\hat{\rm T}\;=\;i\int \bar{Q}(x) S (x,y) 
Q(y) G_{s}(x-y)d^4(x-y)d^4z\;\;.
\label{amplitude}
\ee  
Here $x$ and $y$ are the position of the heavy quarks, and $z$ is the 
instanton center
(the integration over $x+y$ yields the $\de$-function in the 
transition amplitude
expressing the conservation of the total $4$-momentum, which we 
do 
not write explicitly).
$Q(x)$ is the field of a heavy quark with mass $m_Q$ in the 
instanton background, $S (x,y)$
is  the Green function of the massless quark in the instanton 
background, and $G_{s}(x-y)$
represents the product of all color singlet particle (non hadrons) 
Green's functions  
produced in the decay.
Note that all Lorentz indices are suppressed, as well as the 
integration 
over the instanton parameters, other than its center.
For simplicity, we do not explicitly indicate the
dependence of the fields and the quark Green function on the 
instanton collective coordinates, except for position.
In the following, we will use the singular gauge for the instanton 
field. In principle, speaking of instantons assumes that expressions 
are 
written in Euclidean space, but so far the exact nature of the external 
field is 
inessential.
  
We will always assume the heavy hadron is at rest, and thus we can  
single out the large  ``mechanical" part of the $x$-dependence in 
$Q(x)\:$:
\beq
  Q(x)={\rm e}\,^{-im_Qt}\tilde{Q}(x)\;\;.
\label{exp}
\eeq    
Calculating the width we will need to calculate the expectation value 
of
the transition operator between the heavy hadron state:
\beq
T= \frac{1}{2M_{H_Q}}\matel{H_Q}{\hat T}{H_Q}
\label{amp}
\eeq
where now  
\beq
\hat{T}= 
i\,\int \bar{\tilde Q}(x) S (x,y)
\tilde Q(y) G_{s}(x-y)\;{\rm e}\,^{im_Q(x_0-y_0)}\,d^4(x-y)\;d^4z
\;\;.
\label{oper}
\eeq
Since the product of the quark Green functions and nonrelativistic 
$\tilde Q$
fields does not have an explicit strong dependence on $m_Q$, we 
clearly deal with a 
hard (momentum $\sim m_Q$) Fourier transform of a certain 
hadronic correlator 
which is soft in what concerns nonperturbative effects. Note that 
$m_Q$ can
now be considered 
an external parameter in the problem, for example, as an 
arbitrary, and even complex, number. We are not yet formally ready, 
however,  
to consider the Euclidean theory since we still have initial and final 
states. 
We shall address this issue a bit later, and now proceed as if we deal 
with
free heavy quarks, which are transferred to the Euclidean domain 
without problems.  

\begin{figure}
  \epsfxsize=9cm
  \centerline{\epsfbox{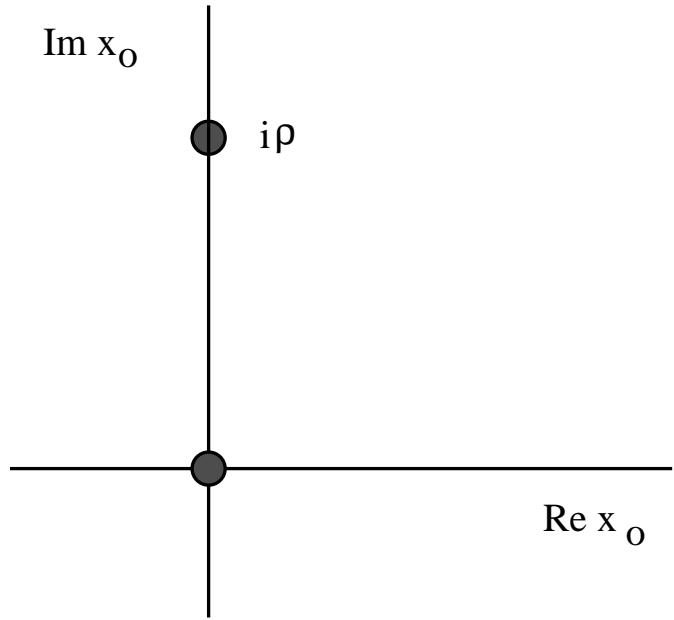}}
  \caption{Finite distance singularity  of the Green function of
  a massless particle in the instanton background. }
  \label{fig4}
\end{figure}

Let us examine the propagator of the massless quark in the  
instanton 
background, which is calculated 
exactly for  the case of spin 0, 1/2, 1 particles 
\cite{Brown}. 
This (Euclidean) Green function $S(x,y)$ has a generic form 
\bb
 S(x,y) =\frac{1}{[(x-y)^2]^n}\frac{1}
{[(x-z)^2 +\rho^2]^{k_1/2}\,[(y-z)^2 +\rho^2]^{k_2/2}}
\times \tilde{S} 
\ee 
where $\tilde{S}$ has no singularities at complex $x_\al$ or $y_\al$ 
(a polynomial).
Using the Feynman parametrization, we rewrite it as 
$$
 S(x,y)\; =\; \frac{\Gamma(k)}{\Gamma(k_1/2)\Gamma(k_2/2)} 
\:
\frac{1}{[(x-y)^2]^n}\;\tilde{S}\; \times
$$
\beq
\int_0^1\, d\xi \:\xi^{k_1/2-1} (1-\xi)^{k_2/2-1}
\;\frac{1}{[\xi(1-\xi)(x-y)^2   + (\rho^2 + \tilde{z}^2)]^k}\;\;,
\label{prop}
\eeq 
$$
k =\frac{k_1+k_2}{2}\;\;,\qquad\; \tilde{z}=z-\xi x-(1-\xi)y\;\;.
$$
For the propagator of a spin $0$ or $1/2$
particle, the value of $k$ is eventually $1$ and $2$, respectively, and 
$n=\;1,\;2$. 
The large-momentum behavior of the Fourier transform of the 
correlator, 
\eq{oper}, depends on the analytic properties of the integrated 
function.
Let us first consider the analytic properties of $S(x,y)$ in the 
complex $(x_0-y_0)$ plane (Fig.\ref{fig4})
  - it has two 
different singularities.
One singularity is on the real axis, and corresponds
to two quarks being at the same point. This is the same singularity 
occuring in the Green function of free 
quarks, but upon integration, the residue is softly modified by the 
instanton field. Picking up this 
pole and calculating the amplitude, we will get instanton 
contributions to the usual power $(1/m_Q\rho)^n$ terms in the OPE, 
which
we are not interested in.
Indeed, making a Taylor expansion
in $(x-y)/\rho$ around this pole, we obtain a series of
corrections $((x-y)/\rho)^k/(x-y)^{2n}$, which, integrated with the 
exponent, result in the above terms.

Another singularity lies on the imaginary $(x_0-y_0)$ axis. It 
comes from the finite quark separation 
\beq
(x-y)^2 = [\xi(1-\xi)]^{-1}(\rho^2 + \tilde{z}^2)\;\;.
\label{sing}
\eeq
In contrast to the \prt or OPE pieces, this separation does not scale
like $1/m_Q$, but stays finite in the heavy quark limit. 
Upon integration over $d^4z\,d^4(x-y)\,d\xi$ this 
singularity,  together with the factor ${\rm e}^{im_Q (x_0-y_0)}$ 
from 
the 
heavy fields, produces the
$
{\rm e}\,^{-{\rm const}\; m_Q\rho}
$
terms in the (Euclidean) amplitude that we are looking for.
We then only need to determine the constant that enters the 
exponent, and the
pre-exponential factor.

Now with this general strategy in mind, let us outline the machinery 
in more
detail. We want to abstract from the complicated questions of the 
interrelation
of the instanton configurations to the particular heavy hadron 
structure,
i.e. to consider the simplest possible state similar to a quasifree 
heavy quark 
instead of a real $B$ or $D$ meson or heavy baryon. On the other 
hand, the
heavy quark,{\em a priori}, cannot be taken as free since the field 
$Q(x)$ must
obey the QCD equation of motion, in particular, in the instanton field. 
It is clear that such a program can be carried out consistently if the
instanton size is small enough compared to the typical size of the 
hadron
$\sim \Lam$, but still is much larger than $m_Q^{-1}$. Having this 
choice  in mind, we  neglect, in what 
follows, the fact that the heavy quark is actually bound in the hadron 
although, eventually, the  values of $\rho$ will not be parametrically 
smaller 
than the  hadronic scale.

Thus we merely solve the equation of motion for the heavy quark in 
the instanton field, as one would do for an isolated particle. The role 
of 
the initial hadronic state, $H_Q$, in \eq{amp} is played by the single 
heavy quark spread in space and evolving in time according to the 
solution 
of the heavy quark Dirac equation analytically continued from 
Euclidean to 
Minkowski space. In our actual calculations we, of course, go in 
reverse:
both the heavy quark field and the transition operator are calculated 
in 
the Euclidean domain; the subsequent continuation to the Minkowski 
space is 
performed in the final expression for the forward {\em amplitude} 
$T$. 
Technically, we are able to solve the equation of
motion for the heavy quark field since the parameter $ m_Q \rho \gg 
1$.

The heavy field $\tilde{Q}(x_0,\vec x)$ can be written in the leading 
order as
$$
\tilde Q(x_0, \vec x) = {\rm T\:e}\,^{i\int_{0}^{x_0}
 A_0(\tau, \vec{x})d\tau} \;\tilde Q(0, \vec x)
 + {\cal O}\left(1/(m_Q\rho) \right) \equiv
$$
\beq
 U(x)\, \tilde Q(0, \vec x) + {\cal O}\left(1/(m_Q\rho) \right)\;\;.
\label{pexp}
\eeq 
The expression is
written in Euclidean space, although we use Minkowski 
notations.
Using the explicit solution for the $SU(2)$ instanton in the 
singular gauge, \eq{inst}, one gets the 
matrices $U$ in the following cumbersome form: 
$$                                       
U(x)= \exp{\left\{i\,\vec\tau \vec n
\left[\left(\arctan{\left(\frac{z_0}{|\vec z -\vec x\,|}\right)}-
\arctan{\left(\frac{z_0-x_0}{|\vec z -\vec x\,|}\right)} \right)-
\right.\right.}
$$
\beq
\left.\left.
-
\frac{|\vec z -\vec x|}{\sqrt{(\vec z -\vec x\,)^2+\rho^2} }
\left(\arctan{\left(\frac{z_0}{\sqrt{(\vec z -\vec 
x\,)^2+\rho^2}}\right)}-
\arctan{\left(\frac{z_0-x_0}{\sqrt{(\vec z -\vec 
x\,)^2+\rho^2}}\right)} 
\right)
\right] \right\} 
\label{u}
\end{equation}
where $\vec n = (\vec{x}-\vec{z})/
\sqrt{(\vec{x}-\vec{z}\,)^{2}}$, $z$ is the coordinate of the
instanton's center 
and 
$\vec\tau/2$ are the color $SU(2)$ generators. The integration is 
simplified since along the integration path $\vec{x}-\vec{z}=\rm 
const$ in 
\eq{pexp}, $A_0$ is proportional to one and the same color matrix 
$\vec\tau(\vec{x}-\vec{z}\,)$ and, therefore, the
path-ordered exponent reduces to the usual exponent of the integral 
of $A_0$.
 
It is important that the expression for $Q(x)$ is only valid in the
leading order in the expansion parameter $1/m_Q\rho$. If one 
considers the contribution of small size instantons (as in 
Ref.~\cite{Chy}), 
then no legitimate expansion parameter is available. The expansion 
in the 
heavy quark mass can only be obtained if instantons of size $\rho < 
\rho_c $ (where 
$\rho_c$ is some parameter $\gg 1/m_Q$) are absent. 
Otherwise, the corresponding equations of 
motion need to be solved exactly.

Even though we managed to solve the equations of motion for the 
heavy quarks in the leading approximation, the solution governed by 
the color 
matrices $U$ is not analytic. The apparent singularities at $x=z$ or 
$y=z$ are,
in fact, spurious and merely an artifact of using the singular gauge 
for the 
instanton field; since the amplitude we calculate is manifestly gauge 
invariant
(it is nothing but the light quark Green functions times the path
exponent~\footnote{Since the path exponent over the straight line is 
unity in the Fock-Schwinger gauge, the products we calculate are 
nothing else 
than the quark Green functions in the instanton field in the Fock-
Schwinger 
gauge. However, the fixed point of the gauge does not lie at the 
center of the instanton, but, rather, at the external current point, $x$ 
or 
$y$. For a review of the Schwinger gauge see Ref. \cite{FSch}.}), 
this singularity is absent in the full expression, being canceled by 
similar terms in the light quark Green functions. However, in 
general, the  
propagation matrix $U$ introduces additional exponential corrections, 
since 
$$
\arctan{\left(\frac{t}{\sqrt{\vec{x}^{\,2} +\rho^{\,2}}}\right)}
$$
has a (cut) singularity at 
$t=i\sqrt{\vec{x}^{\,2} +\rho^2}$, which is a point where
the singularity in the potential is not of the gauge type.
The fortunate simplification which arises, in the leading 
approximation in $1/(m_Q\rho)$, is that the two factors $U(y)$ and 
$U^{-1}(x)$ 
are unity at the saddle point. This happens due to the fact that the 
saddle point 
corresponds to the configuration where the instanton is situated right 
on the line (in three-dimensional coordinate space) connecting 
$\vec x$ and $\vec y$ and thus $\vec x - \vec z = \vec y - \vec z = 
0$ \eq{sadd}
Finally, picking up the pole in the complex $(x_0-y_0)$ plane  in the 
light quark propagator in 
\eq{prop},
\beq
[((x-y)^2 + [\xi(1-\xi)]^{-1} (\rho^2 + \tilde{z}^2)]^{-k}\, ,
\label{pole}
\eeq
we get an exponential factor
$$
\exp{\left(-m_Q\sqrt{(\rho^2 +\tilde z^2)/(\xi(1-\xi))  
+ (\vec{x}-\vec{y}\,)^2}\,\right)}
$$ 
in the transition amplitude. The expression in the exponent has a 
sharp 
minimum at
\bb
\tilde z=0,\quad \xi=1/2, \quad (\vec{x}-\vec{y})^2=0 \;\;.
\label{sadd}
\ee 
Evaluating it at this point we get the exponential factor
$$
{\rm e}\,^{-2m_Q \rho}\;\;.
$$

The power of $m_Q$ in the pre exponent can be determined without 
actual calculations as well. The residue of the $k$-th order pole of 
the 
propagator yields the factor $m_Q^{k-1}$ upon integration over $x_0-
y_0$. The 
Gaussian integrals over $(\vec{x}-\vec{y})$, $x_0-y_0$, $\tilde{z}$, 
and $\xi$ 
around their saddle points give $m_Q^{-4}$; on the contrary, 
all ``free'' propagators 
(those of the color singlet final particles and the bare propagators of 
the 
quarks produced) enter at a fixed 
separation $\sim
-\rho^2$, and are $m_Q$-independent. We thus get
\beq
T \; \propto\; {\rm const} \cdot m_Q^{k-5}\,{\rm e}\,^{-2m_Q 
\rho}\;\;. 
\label{power}
\eeq
For example, the large size instanton corrections to the transition 
amplitude
for the semileptonic decays of heavy quarks has the form 
\beq
T_{\rm sl} \; \propto\;\frac{{\rm e}\,^{-2m_Q \rho}}
{m_Q^3}
\;\;.
\label{pow}
\eeq
The pre exponent can be easily calculated in the same 
way and will be given for a few cases of interest in the subsequent  
sections.

\begin{figure}
  \epsfxsize=12cm
  \centerline{\epsfbox{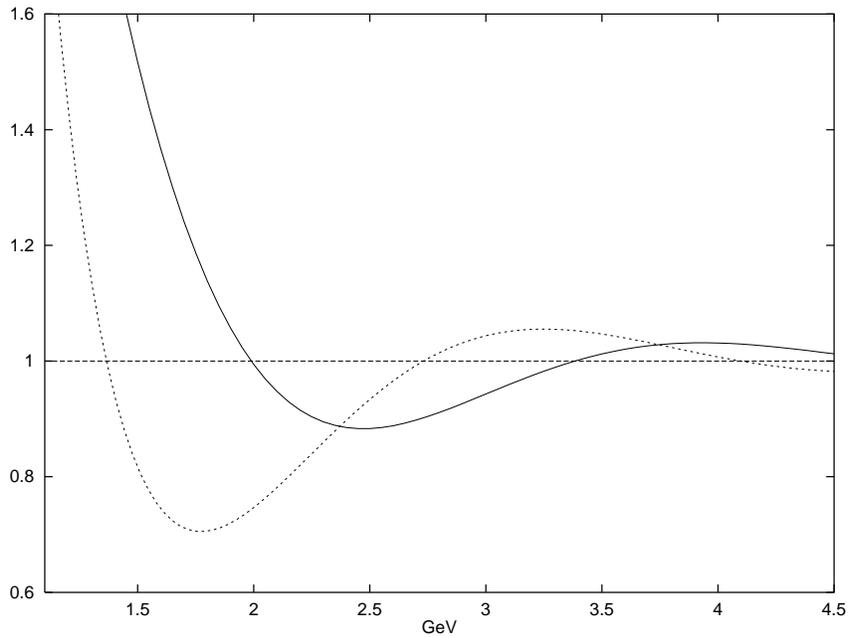}}
  \caption{Exact (solid line) and asymptotic (dashed line)
                       behavior of $R(E)$.  }
  \label{fig5}
\end{figure}

The same counting rules apply for the case when more quarks are 
present in the
final state. In the case of the vector correlator of the light quarks we 
have 
the product of two light quark propagators instead of one, and 
eventually 
$k=4$. Since
$\Pi_{\mu\nu}(Q) \sim Q^2 \Pi(Q^2)$, we get for $\Pi(Q^2)$
defined in \eq{b2}
\bb
\Pi(Q^2) \;=\;4\pi^3  \;\frac{{\rm e}\,^{-
2\sqrt{Q^2}\rho}}{(\;Q\;\rho\;)^3}\;\;,
\ee
in accordance with the explicit calculation of \eq{b2}.
In fact, the asymptotic expression works accurately enough even in 
the 
Minkowski domain already at $\sqrt{s} \rho \simeq 3$; the 
corresponding approximate 
expression for $\frac{1}{\pi}\Im \Pi$ is plotted as a dashed curve in 
Fig.\ref{fig5}.
It is clear that the main effect of the subleading in $1/(Q\rho)$ terms 
is a
phase shift in the oscillations.

Using the same counting techniques, one concludes that the non-OPE 
soft instanton 
contribution to the forward amplitude describing nonleptonic decays 
of the heavy 
quark (we have three light quarks in the final state i.e. $k = 6$)
scales like  
\beq
T_{\rm nl} \; \propto\; m_Q \,{\rm e}\,^{-2m_Q \rho}
\;\;.
\label{pownl}
\eeq
In the next section, we give a more detailed calculation, and 
discuss total widths and differential distributions.

A qualifying comment is in order here. So far, in calculating the 
exponential terms, we assumed the heavy initial quark to be static
(at rest). In this approximation the presence of the initial heavy 
quark does not affect our result at all; it is the final quarks that fully 
determine the exponential terms.  We know for sure, however, that 
the initial heavy quark experiences a ``Fermi" motion inside the 
heavy 
hadron. In the practical OPE, the first correction due to this Fermi motion  
comes from the operator $\bar Q {\vec\pi}^2 Q$. 
In other words, the accuracy of the static initial quark approximation
is $1/m_Q^2$. An important question is
how the exponential terms of the type we focus on show up in those 
subleading effects proportional to $\langle {\vec\pi}^2\rangle$ (and 
other similar 
subleading effects).

The expansion of the heavy quark propagator 
in $1/m_Q$ leads,  generally speaking, to terms of the type 
$(1/m_Q^2)1/(x^2+\rho^2)$, 
i.e. we get an extra singularity in the amplitude due to the initial 
heavy quark. This extra factor enhances the overall singularity of the 
amplitude in the complex plane and, thus, leads to a higher power of 
$m_Q$ in the preexponent. The additional denominator, 
$(x^2+\rho^2)$, 
generates only the first power of $m_Q$, however, so the overall 
effect is
suppressed by the small parameter $1/(m_Q\rho)$.

In the Minkowski domain, inside the oscillation zone, the exponential 
factor is not a suppression at all, so we must count only the 
pre-exponential factors. We see that deviations from duality
are parametrically relatively stronger in the $\langle 
{\vec\pi}^2\rangle$
piece. Still the original suppression of the $\langle 
{\vec\pi}^2\rangle$ piece by $1/m_Q^2$ is not completely lifted. 
We lose one power of $1/m_Q$, but retain the other power. 
Thus, our approximation of the static initial quark in the analysis
of the exponential deviations from duality is justified. 
Nevertheless, it is interesting 
to note that the initial-state $1/m_Q$ effects are less suppressed in 
the exponential terms. This  seems to be a  general 
feature. 

\subsection{Decays into massive quarks}

The case of the massive final state quark (e.g. $b\rightarrow c 
\l\nu$) does not differ conceptually 
if treated in the $1/(Q\rho)$ expansion, although a technical 
complication arises due to the unknown explicit expression for the 
massive 
propagator in the instanton field. Still, this is not a stumbling block in 
the 
analysis: the relevant 
singularity of the $c$ quark Green function can again only be at 
  $(x-z)^2=-\rho^2$ or $(y-z)^2=-\rho^2$, 
and the corresponding power of the displacements can be 
determined analytically keeping trace of the singular terms in the 
massive 
Dirac equation. Only the exact constant in front of this 
singular term constitutes a problem, and it can be evaluated 
numerically. We 
shall address this case in detail elsewhere, and here only consider a 
few 
limiting cases.

\vspace{0.3cm}

{\it (i) Heavy and light quarks in the final state, e.g.} $b\rightarrow
c d\bar u$

\vspace{0.1cm}
 
It is possible to see that, in this case, the exponential terms,
in the leading approximation, are associated with the light quark, and 
the presence of the heavy quark in the final state has no impact 
apart from changing kinematics. Indeed, 
we saw that the Green functions of the final state particles enter at 
large
distances $|x| \sim \rho$ (it will be also illustrated in more detail in
the next 
section). In this situation 
the interaction of the final 
heavy quark with the soft background field reduces to the 
ordered exponential of $i\int A_\mu (\xi ) d\xi_\mu $ along the 
heavy 
quark trajectory. 
We have already calculated it and found to be unity (the exponent
to vanish) in the saddle point configuration.
Therefore the final heavy 
quark can be taken as non-interacting. The propagator of the 
non-interacting final $c$ quark $\propto {\rm e}^{-im_c |x|}$
(${\rm e}^{-m_c |x|}$ in the
Euclidean time) will be multiplied by the light quark Green 
function,
which develops a pole in the complex $x^2$ plane.  This means that 
as far as the exponential terms are concerned, the effect of the final 
heavy quark is the replacement of $m_Q$ in the Fourier transform 
by
$m_Q - m_c$, 
\beq
m_Q\;\ra \Delta\;=\;m_Q-m_{\rm fin}
\label{f7}
\eeq
(in the general case, with a few heavy particles in the final state, 
$\Delta=m_Q-\sum m_{\rm fin}\,$, with the sum running over all 
heavy final state
particles). Calculating the width, one thus has, at the saddle point, the 
exponent of the form ${\rm e}^{-2\Delta\rho}$ rather than ${\rm 
e}^{-2m_Q 
\rho}$. It is worth emphasizing that this result holds as long as the 
mass of 
the final state quark involved exceeds $1/\rho$, i.e.  $m_{\rm 
fin}\,\rho \gg 
1$, 
regardless of the actual hierarchy between $m_Q$ and $m_{\rm 
fin}$.
The very same kinematic change refers, of course, also to the case 
when massive leptons (or other color-singlet particles) are present in 
the final state.

Such a result might seem counter-intuitive to the reader who 
would compare it, say,  with the free quark answer for the total 
width, where  
the final quark mass appears only as a quadratic correction $\propto 
m_{\rm 
fin}^2/m_Q^2$ if  $m_{\rm fin}/m_Q \ll 1$, and the final heavy  
quark 
is typically fast. As a matter of fact, there is no mystery -- the 
occurrence of 
${\rm e}^{-2\Delta\rho}$, instead of ${\rm e}^{-2m_Q \rho}$,
is a manifestation of a remarkable property of the exponential terms 
discussed in Sect. 5. These terms are determined by the highest 
possible invariant mass of the  light quark system.  In the case at 
hand this is achieved when the final $c$ quark is at rest, i.e. we are 
in the SV limit  \cite{SVlimit}. With this picture, it is then no surprise 
that
the exponential terms are determined by ${\rm e}^{-2\Delta\rho}$.
 
Since this result is rather unusual, we reiterate.
In calculating the exponential terms in the transitions of the type 
$b\rightarrow c \bar u d$ we find ourselves in the situation of the 
heavy quark  symmetry \cite{SVlimit,IW}. It is an extended 
symmetry, however,
since it applies to {\em arbitrary}  color structure of the weak
vertices, e.g. even when  color flows, say, from $Q$ to the light 
quark $q$ rather than from $Q$ to  the heavy final quark $c$ 
(the standard heavy quark symmetry works only for the color 
singlet $\bar b c$ 
currents). 
Similar consideration applies even to the decays like 
$b\rightarrow c \bar c s$.

\vspace{0.3cm}

{\it (ii) SV limit.}

\vspace{0.1cm}

In the SV limit, when the final heavy quark is slow, we can calculate 
the exponential terms analytically in a wider class of processes, e.g. 
in the nonleptonic transition $b\ra c u d$. 

This case can be treated as follows.  The 
final heavy quark Green function is given  by the $P$ exponent 
together 
with the ``mechanical'' phase factor (we write it here in the Euclidean 
space):
$$
G^{\rm SV}(x,0)\;= \; m^3\frac{1+i\frac{x\gamma}{|x|}}{2}
\,\frac{{\rm e}^{-m|x |}}{(2\pi m|x|)^{3/2}} \,
{\rm T\:e}\,^{i\int_{0}^{x}
 A_\mu(\xi)d\xi_\mu} \;\simeq
$$
\beq
\simeq \;
m^3\frac{1+i\frac{x\gamma}{|x|}}{2}
\,\frac{{\rm e}^{-m|x|}}{(2\pi m|x|)^{3/2}} \,
U^+(x)U(0)\; .
\label{f9}
\eeq
The path exponent is again given by the $U$ matrix presented in 
\eq{u}~\footnote{Here, we can first deform the integration contour 
over $x_0$ into the complex plane and then use the $1/m_{\rm fin}$
expansion for the quark propagator. Another  clarifying remark: for 
massive particles, the propagator entering at the
saddle (effectively Minkowski) point is complex. In doing the saddle 
point 
calculations, one should 
take its values at the ``bottom'' of the cut; we shall dwell on this
point elsewhere.}.  
At the saddle point when the
instanton is on the line $(\vec x,\vec y\,)$ the final heavy quark 
path
exponent equals to unity, and,  therefore, the heavy quarks appear to 
be
``sterile" -- they do not feel  the instanton  field at all. 
Thus, in nonleptonic decays in the SV limit 
the appearance of  
exponential corrections is due - in the leading order - only to the 
light quark interaction with the instanton.
(Note also, in the semileptonic decay in the SV limit the exponential 
terms are absent in the approximation accepted in the present paper. 
To see them we have to go beyond the leading approximation near 
the saddle point approximation. )

It is worth  noting that,
outside  the exponential factor, in the pre-exponent, the masses enter
in a more complicated way than is indicated in Eq. (\ref{f7}). 
For example, the  integration over $d^3 x$ near
the saddle point produces $m_Q^{-3/2}$ rather 
than the power of the energy release.

\subsection{Summary}

Now we are ready to incorporate the effect of both massive and 
massless quarks in the final state.
 
1) Each light (massless) quark contains a second order pole in 
its propagator (\ref{prop}) and brings in two powers of energy 
release in the numerator.

2) Each heavy (static) quark propagator contains factor 
$m^{3/2}_{\rm fin}$

3) The scale of energy release in the exponent is set up by the 
masses of the heavy quarks, 
$$
m_{Q}\; - \;\sum m_{\rm fin}
$$
Exploiting  \eq{power} and discussion above it  we obtain 
\beq
T\;  \propto \;{\rm const} \cdot \left(m_{Q }  -  \sum m_{\rm 
fin}\right)^{2n_l-7/2}\;
m_Q^{-3/2}\;
{\rm e}^{-2\rho (m_{Q }-\sum m_{\rm fin})}
\;\prod m_{\rm fin}^{\frac{3}{2}} 
\label{power1}
\eeq
where $n_l $ is the number of massless quarks;  $m_{Q}$ 
is the mass of the initial quark and 
$m_{\rm fin}$ are the masses of the final heavy particles (both 
quarks
and leptons). If there are no heavy particles 
in the final state we return to \eq{power}.

\section {Heavy quarks -- a toy model}
Before proceeding to the actual calculation of the 
instanton contribution to heavy quark decays, we 
will first
perform the same analysis in the simple toy model of scalar 
chromodynamics. 
Without any loss of physical content, we can simplify our
consideration by
neglecting the spin degrees of freedom, as was suggested in 
Ref. \cite{motion}. We also assume the $SU(2)$ color group here.
  
The weak Lagrangian,
\beq
L_W\;=\;h\:Q\bar{q}\,\phi + h.c.,
\eeq
describes the decay of a heavy scalar quark, Q, into a light (massless) 
quark, $q$,,
and a scalar ``photon'', $\phi\,$; the coupling, $h$, has 
dimensions of 
mass.
Both quarks are in the spinor representation of the color group.
 
The basic strategy of our semiclassical calculation of the transition 
amplitude
has been outlined in the previous section; here we work out details.
We calculate the transition operator in Euclidean space using the  
semiclassical approximation, considering 
scalar quarks in the background of an instanton field. Upon returning 
to the Minkowski space, $T$  will acquire an
imaginary part related to the instanton correction to the total decay 
width.

\subsection{Inclusive width}

Consider the transition amplitude,
\beq
T(r)\;=\; \frac{1}{2M_{H_Q}}\matel{H_Q}{\hat T(r)}{H_Q}=
\frac{1}{2M_{H_Q}}\matel{H_Q}{\,\int\;d^4x\, {\rm e}\,^{-irx} \;
iT\{L_W(x)\,L_W(0)\} }{H_Q}
\label{amp1}
\eeq
where $r$ is an arbitrary external momentum. Denote 
 the 
$4$-velocity of the heavy hadron by $v_\mu$ ($\vec v =0$). 
Proceeding to the 
nonrelativistic fields, $\tilde Q$, and using \eq{amplitude}, we have 
\beq
\hat{T}(r)\;=\;
\;\;i\,\int \bar{\tilde Q}(x) S (x,0)
\tilde Q(0) G_{\phi}(x^2)\;{\rm e}\,^{i(m_Qv-r)x}\,d^4x\;d^4z
\, ,
\label{operq}
\eeq
where $G_{\phi}$ is the propagator of the scalar photon, and $S(x,y)$ 
is
the propagator of the massless scalar quark in the external 
(instanton)
field. We put $y=0$. 
Addressing the total width, we will only consider $\vec r=0$ and 
$r_0$ 
arbitrary
(and complex); the more general case is relevant for differential 
distributions. 
For heavy quarks the transition amplitude depends only on the
combination $m_Q-r_0\,$: 
$$
\;T(r)={\rm T}(m_Q-r_0)\, .
$$
Choosing an appropriate $r_0$, we select deep Euclidean
kinematics and calculate the amplitude in the presence of an 
instanton.
To this end we write
\beq
m_Q-r_0=ik_0\;\;,\;\;\; (x_0,z_0)\;\ra \; -i(x_0,z_0),
\label{euclid}
\eeq
and
\beq
\hat{{\rm T}}(k_0)\;=\;
\int \bar{\tilde Q}(x) S (x,0)
\tilde Q(0)\; G_{\phi}(x)\;{\rm e}\,^{ikx}\,d^4x\;d^4z,
\label{opere}
\eeq
where now everything is in Euclidean space; in what 
follows it is
assumed that $k_0\sim m_Q$.

The propagator of the massless scalar particle in the instanton
background is known exactly, and in the singular gauge takes the 
following
form \cite{Brown}
\beq
S(x,y )\;=\;
\frac{1}{4\pi^2 (x-y)^2}(1+ \rho^2/ x^2)^{-1/2}
\Biggl(1 +\frac{\rho^2 (\tau^+ x)(\tau y)}{x^2 y^2}\Biggr)
\Bigl(1+ \rho^2/y ^2\Bigr)^{-1/2}
\label{scalar}
\eeq
where, to simplify the expression, we have
assumed that the instanton is 
centered at $z=0$ and lies in a 
particular $SU(2)$ color subgroup,
\beq
A_{\mu a}(x) =\frac{ 2\eta_{\mu\nu a}
(x-z)_{\nu}\rho^2}{(x-z)^2((x-z)^2 + \rho^2)}
\label{inst}
\eeq
and  
\beq
\tau = ( \vec\tau, i)\qquad \tau^+ = ( \vec\tau, -i)\qquad
\tau_{\alpha}^+\tau_{\beta}=\delta_{\alpha\beta}+\eta_{\alpha\beta
c}\tau_{c}\;\;,
\eeq
where $\vec \tau$ are the Pauli matrices acting in the color 
subgroup.

The heavy quark field, $\tilde Q(x)$, is the solution of the 
equation of motion in the instanton background 
$$
iD_0 \tilde Q(x) = \frac{1}{2m_Q} (iD)^2 \tilde Q(x)
$$
which, in the leading order in $1/m_Q$, yields \eq{pexp} with the 
matrices $U$ given by \eq{u}. For the heavy scalar quarks one has
\beq
\frac{1}{2M_{H_Q}}\langle H_Q\mid\bar{Q}Q\mid H_Q\rangle \;=
\;\frac{1}{2m_Q}\;\;.
\label{normscal}
\eeq

Now we take a closer look at the final state quark propagator,
rewriting it using the Feynman parametrization:
$$
S (x,y)\; =\;\frac{1}{\pi} \,\int_0^1d\xi\: 
\left[\xi(1-\xi)\right]^{-1/2}\;
\frac{1}{ \xi(1-\xi)(x-y)^2+ \rho^2+\tilde{z}^2  }\times
$$
\beq
\frac{1}{4\pi^2 (x-y)^2}\;
\Biggl(1 +\frac{\rho^2 \tau^+(x- z)\tau (y-z)}{(x-z)^2 (y-z) ^2}\Biggr)
\; \sqrt{(x-z)^2}\sqrt{(y-z)^2}
\label{props}
\eeq 
where 
$$
\tilde{z}=z-x\xi-(1-\xi)y\;\;.
$$   
Integrating in \eq{opere} over $x_0 $ we only pick up the pole at
$$
x^2= \frac{\rho^2+\tilde{z}^2}{\xi(1-\xi)}
$$
which yields the following exponential factor:
\beq
{\rm e}\,^{-k_0\sqrt{(\rho^2 +\tilde{z}^2)/(\xi(1-\xi)) + \vec{x}^2} 
}\;\;.
\label{saddle}
\eeq
At $k_0\rho\sim m_Q\rho \gg 1$, the remaining integrations are 
nearly 
Gaussian, and run over narrow intervals, 
\beq
\vec{x}^2 \sim \frac{\rho}{m_Q}\;\;,\;\;\;
\left(\xi-\frac{1}{2}\right)^2
\sim \frac{1}{m_Q\rho}\;\;,\;\;\;\left(z-\frac{x}{2}\right)^2 \sim
\frac{\rho}{m_Q}\; .
\label{perev}
\eeq
Thus,  one performs the remaining integrations
by merely evaluating all pre exponential 
factors at the
saddle point. In particular, this  refers to 
the matrix $U^{-1}(x)$, 
coming from the heavy quark propagation, which is now the path 
exponent from
the point $(x_0, 0\,)$ to $(0,0 \,)\,$.
$U^{-1}(x)$ 
evaluated at the saddle
point and is just the  unit matrix,
and    
the color part of the light quark Green function 
\eq{scalar} . The heavy quark field, $\bar{\tilde Q}(0,\vec 
x\,)$,
according to \eq{perev}, enters at  distances $\vec x\sim 
\sqrt{\rho/m_Q}
\ll \rho$ and, therefore, the transition operator is
finally 
proportional 
to 
$\bar Q(0)\,Q(0)\,$.

Collecting all remaining factors, one has
\beq
\hat{{\rm T}}(k_0)\; = \;h^2\; \bar Q(0)\left\{ 
\frac{G_\phi(-4\rho^2)}{2 \pi^{2}\rho}\;
\int\; d\xi\:d^3\vec{x} \:d^4 z\;
{\rm e}\,^{-k_0\sqrt{(\rho^2 +z^2)/(\xi(1-\xi)) + \vec{x}^2)}}
\right\} Q(0) \;\;,
\label{transcal}
\eeq
where 
 $$
G_\phi(x^2)=\frac{1}{4\pi^2 x^2},
$$
is the free scalar propagator.
Performing the Gaussian integrations (which yield the factor
$4\pi^4\rho^3k_0^{-4}\cdot {\rm e}^{-2k_0\rho}$),
and using \eq{normscal}, we finally arrive at
\beq
T(q_0)\; =\; -\frac{h^2}{16m_Q}\; \frac{{\rm e}\,^{-
2k_0\rho}}{k_0^4}\;=\;-
h^2\; \frac{{\rm e}\,^{2i(m_Q-r_0)\rho}}{16m_Q(m_Q-r_0)^4}
\;\;.
\label{scalfin}
\eeq
 Let us 
note that the above equation shows 
the correct pre-exponential power of $m_Q$. The propagator of the 
scalar particle in the instanton background 
has a first order pole, unlike the second order one of the spin-half 
quark,  and matrix element (\ref{normscal}) has an additional power 
of 
$m_Q^{-1}$.
This leads to a result with two powers of $m_Q\rho$ less then 
predicted 
in \eq{pow}.

Now, we are interested in the decay width, which is given by the 
imaginary part
of $T(r)$ at the physical Minkowski point $r=0$. Since the 
singularity of the
amplitude we calculated is located far enough away, at $r_0\simeq 
m_Q$, 
we perform 
a straightforward analytic continuation by merely setting $r_0=0$ 
and, thus, get
\beq
\Gamma^{\rm I}_{\rm scal}\;=\; - h^2\;  \;  
 \; \frac{\sin{(2m_Q\rho)}}{8m_Q^5\;\rho^4}\;\;.
\label{scalwidth}
\eeq
In accordance with the general analysis of Sect.~3 the instanton 
contribution in the width 
decreases only as 
a power of $m_Q$ and oscillates.

\subsection{Decay distribution}

Here we address the calculation of the instanton-induced corrections 
to the 
``photon'' spectrum. The corrections blow up near the tree-level 
endpoint, where
the semiclassical approach is not applicable. On the other hand, the 
total
width is given by the integral over the whole spectrum and is 
calculable. The
situation here is similar to the one we discussed in the Sect.~5 for the 
finite
energy moment integrals in the
$\rm e^+e^-$ cross section, and we  
closely
follow this analogy in our analysis.

Let us denote the photon momentum by $q$, with $q_0=E_\ga$. To 
find $d\Gamma/dE_\ga$, one must consider the transition
operator similar to the total width,
but without the photon propagator $G_\phi\,$:
\beq
dT(r,q)\;=\; \frac{1}{2M_{H_Q}}\matel{H_Q}{d \hat T(q,r)}{H_Q}
\label{damp}
\eeq
with
 \beq
d\hat T(q,r)\;=\;
\;i \,\int \bar{\tilde Q}(x) S (x,0)
\tilde Q(0)\;{\rm e}\,^{i(m_Qv-q-r)x}\,d^4x\;d^4z
\;\;.
\label{doper}
\eeq
The transition operator then  clearly depends only on the sum of the 
four-momenta $q+r$, but 
since
calculating the differential decay rate we need to keep $q^2=0$ and
$q_0=E_\ga >0$, such a temporary proliferation formally allows us to 
use the
momentum $r$ for analytic continuation and is appropriate. In 
particular, we
will keep $\vec{q}$ fixed, say, directed along the $z$ axis, assume 
$\vec
r=0$, and again use the variable $r_0$ to make calculations in the 
Euclidean
domain. Kinematically, the amplitude $dT(r,q)$ depends on two 
invariants,
$(r+q)^2$ and $(r+q)_0$, or any two combinations thereof.

Technically, the calculation of the transition operator does not differ 
from
the case of the total width except for the fact that now the Euclidean 
momentum $k$
$$
m_Q-q_0-r_0=ik_0\;\;\;,\;\;\;\; \vec{q}+\vec{r}=\vec k
$$
has not only the zeroth, but also spacelike components. The saddle 
point calculation goes exactly in the same way if one
 replaces $k_0$ by
$\sqrt{k^2}=\sqrt{k_0^2 + \vec k^{\,2} }$, the main contribution 
comes from the
singularity of the light quark propagator and the heavy quark 
matrix $U$ still equals unity at the saddle point. 
For this reason, in the leading approximation the
amplitude $dT(r,q)$ appears to depend, in fact,  only on one 
kinematic 
variable 
$(m_Qv-r-q)^2=(p-q)^2$, and one has
\beq
dT(q,r)\; =\; \pi^2\frac{h^2\rho^2}{m_Q}\; 
\frac{{\rm e}\,^{-2\rho\sqrt{k^2}}}{ k^4}\;=\; 
\pi^2\; \frac{h^2\rho^2}{m_Q}\;
\frac{{\rm e}\,^{2i\rho\sqrt{(m_Qv-q)^2}}}{(m_Qv-q)^4\;\rho^4}
\;\;.   
\label{dscalfin}
\eeq
In the last equation, we continued the result to the physical domain 
setting $r=0$;
to be far enough from the singularity and ensure the applicability of 
the calculations we must assume that $(m_Qv-q)^2 \rho^2 \gg 1$.

The differential decay rate for the massless $\phi$ is given by
$$
\frac{d\Gamma}{dq_0}=  \frac{1}{2\pi^2}  q_0 \,\vartheta(q_0)\;
\Im dT(q) \;= 
$$
\beq
=\; \frac{h^2\rho^2}{2m_Q}  q_0 
\,\vartheta(q_0)\cdot\vartheta(m_Q-
2q_0)
 \frac{\sin{\left(2\rho\sqrt{(m_Qv-q)^2}\; \right)}} {(m_Qv-
q)^4\;\rho^4} 
\label{spectrum}
\eeq
(remember, we fixed $|\vec q\,|=q_0$).
\Eq{dscalfin} (and (\ref{spectrum}) for the imaginary part), in fact, 
are 
nothing but the soft instanton
contribution to the massless (scalar) quark propagator (in the 
Fock-Schwinger gauge, or dressed with 
the path exponent to make it gauge-invariant in the general case).

In the approximation of free quark decay, the photon spectrum is 
monochromatic, $\propto
\de(E_\ga-m_Q/2)$. Equation (\ref{spectrum}) 
yields a decay 
spectrum below the end 
point, at $E_{\gamma} < (m_Q/2)$. The result for the spectrum
is, as expected, oscillating (sign-alternating). This does not lead to 
any
physical problems, of course, because  this contribution is to 
be
considered on the background of ``normal'' OPE corrections (as well 
as the
perturbative ones) which populate the spectrum below the 
two-body
endpoint. The main OPE contribution near the end point is related to 
certain 
initial-state
interactions of the heavy quark, and is interpreted as  
Fermi 
motion \cite{prl,motion,Neubert}. It produces a decay 
distribution 
which is unsuppressed, of 
order
unity, in the interval $|E_\ga-m_Q/2| \sim \Lam$, and decreases fast 
only when 
$m_Q/2-E_\ga$ becomes larger than a hadronic scale. 
At $(m_Q/2) -E_{\gamma} \gg \Lam$ the perturbative tail of the 
spectrum density takes over. 
In this domain 
our
instanton calculations are already legitimate, since there 
$\sqrt{(m_Qv-q)^2}
\sim \sqrt{m_Q\rho}$ if $\rho \sim \Lam$. However, we emphasize 
that the
corrections (\ref{dscalfin}), (\ref{spectrum}) are {\em not} related to 
the Fermi
motion: the instanton effects in the latter appear in  subleading 
orders in
$m_Q\rho$.

Equation (\ref{spectrum}), far enough from the end point, represents, 
in a sense, 
a purely
``exponential'' effect. This is not the case anymore, however, if one 
attempts to
integrate  over  $E_\ga$ and find the contribution to the 
total
decay width. This fact is clearly revealed at the technical level when 
one 
compares
the above calculation with the preceeding calculation of
the total width: in 
the case
of $\Gamma_{\rm tot}$ one {\it effectively} uses the 
$x$-independent 
photon
propagator $G_\phi(x^2)=-1/(16\pi^2\rho^2)$, whereas 
\eq{spectrum}, 
according to 
Cutkowski's rules, corresponds to the propagator $G_\phi(q)=-1/q^2$, 
i.e. $G_\phi(x^2)=1/(4\pi^2 x^2)$. They differ 
explicitly by the presence of the 
singularity
at $x=0$, which generates the normal OPE terms in $\Gamma_{\rm 
tot}$. Of course,
a thoughtful  instanton calculation of the integrated spectrum yields 
the same result as the direct calculation of the
 total 
width.
Starting from the differential spectrum one has to resort to a special
treatment of the OPE domain near the end point.

A straightforward attempt to calculate the total width by integrating 
the
spectrum (\ref{spectrum}) seemingly faces a surprising problem: the 
correction
grows fast toward the end point 
\cite{Chy,AFalk}
where the light quark is soft (though 
carrying
large energy), and the integral seems to saturate at 
$m_Q/2-E_\ga 
\sim 1/(m_Q\rho^2)$, where the expansion in $1/Q\rho$ fails, and 
the 
overall result
completely depends on an {\em ad hoc} cutoff procedure; the total 
contribution
then would allegedly be governed by this soft scale, rather than by 
$m_Q$. This 
is similar
to the apparent paradox in the finite energy moments of $R$ in the 
${\rm
e^+e^-}$ annihilation addressed in Sect.~5. The resolution of the 
paradox is 
essentially the
same: any exponential contribution in the hard part of the spectrum 
is
automatically accompanied by the corresponding OPE-like 
``counterterms'' in the
end point region, which are superficially invisible. Although the exact 
spectrum in the latter cannot be
calculated, certain integrals are defined unambiguously. As a result, 
the
non-zero total width effect emerges only from the explicit 
(non analytic)
constraint on the photon energy   at $E_\ga=0$, i.e. at {\em 
small} $q$,
and, thus, the instanton (exponential) contribution to the total width 
is indeed determined by the hard scale 
$m_Q$.

To reveal this conspiracy explicitly, we again exploit the fact that 
$d\hat
T$ exponentially decreases in the Euclidean domain. In our concrete 
case, there
is a technical simplification:  $d\hat T(q)$ depends only on 
one
kinematic variable $\kappa^2=(m_Qv-q)^2$, the ``invariant mass 
squared'' of the
light final state quark. One can write the dispersion relation over 
$\kappa^2$,
\beq
dT(\kappa^2)\;=\; \frac{1}{\pi}\,\int_0^\infty \frac{\Im
dT(\kappa'^2)}{\kappa'^2-\kappa^2} \;d\kappa'^2\;  .
\label{c2}
\eeq
The fact that  $dT(\kappa^2)$ falls off exponentially in the Euclidean 
domain, 
$dT(\kappa^2) \sim
{\rm e}^{-2\sqrt{-\kappa^2}\rho}$, means that all moments of $\Im 
dT(\kappa^2)$
vanish:
\beq
\int_0^\infty\; \kappa^{2n}\,\Im dT(\kappa^2)
\;d\kappa^2\;=\;0\;\;.
\label{c3}
\eeq
On the other hand, $\Im dT(\kappa^2)$ differs from 
$d\Gamma/dq_0$ 
by only a simple kinematic factor. For a massless $\phi$, for 
example,
\beq
\frac{d\Gamma}{dq_0}\;=\;\frac{1}{2\pi^2} \;q_0 \,\vartheta(q_0)\:
\Im dT(q)\;\;,
\label{c4}
\eeq
$$
q_0=\frac{m_Q^2-\kappa^2}{2m_Q}\;\;, \qquad |\vec q\,|=q_0\;\;.
$$
Therefore, 
\beq
\int_E^{M_{H_Q}/2}\,dq_0\;q_0^n\,\frac{d\Gamma}{dq_0}\;=\;-
\frac{1}{4\pi^2m_Q}
\,\int_{m_Q^2-
2m_QE}^{\infty}
\; \left(\frac{m_Q^2-\kappa^2}{2m_Q}\right)^{n+1}\Im 
dT(\kappa^2)\;d\kappa^2\; .
\label{c5}
\eeq
We see that the instanton correction to the spectrum integrated from 
the very
end point down to some $E_{\rm min}$ is determined, not by the end 
point effects,
but by the lowest energy included, and one has in this case $2\rho Q 
\ra
2\rho \sqrt{m_Q^2-2m_QE_{\rm min}}\,$. If one integrates over the 
whole spectrum,
the correction is parametrically minimal, and reproduces  the 
correction for the total 
width.
Clearly, it is the most general feature of the ``exponential'' effects 
which
does not depend on the particular decay.
As was mentioned in Sect. 5, the constraints (\ref{c3}) can be written 
in the following compact 
form  
\beq
\Im dT(\kappa^2)\;=\;\int_0^{\infty} \; ds\; \chi(s) \left[\de(s-
\kappa^2)-
{\rm e}\,
^{-s\frac{\partial}{\partial \kappa^2}} \de(\kappa^2) \right],
\label{c6}
\eeq
where $\chi(s)$ is some smooth function vanishing at $s\le 0$. This
expression automatically satisfies the relation
\beq
\int_0^{\kappa^2}\;\Im dT(\kappa^2)\, P(\kappa^2) 
\;d\kappa^2\;=\;-
\int_{\kappa^2}^{\infty}
\; \Im dT(\kappa^2)\, P(\kappa^2)\;d\kappa^2
\label{c7}
\eeq
for any appropriate analytic $P(\kappa^2)$. The saddle point 
instanton
calculation carried out  above determins the asymptotics of the 
function 
$\chi(s)$ at
$s\gg 1/\rho^2\,$:
\beq
\chi(s)\; \simeq \; \pi^2 h^2 \frac{\rho^2}{m_Q}\:
\frac{\sin{(2\sqrt{s}\,\rho)}}{s^2}\;  .
\label{c8}
\eeq
In such a saddle point way we cannot calculate $\Im dT(\kappa^2)$ 
at 
$\kappa \rho
\lsim 1$. According to \eq{c6}, however, we know certain integrals 
over the
small virtuality domain. For any particular ``exact" $\chi(s)$ they are 
obtained
as a series of $\de$ functions and derivatives of $\de$ functions
located at the end point, 
which, when summed, give a certain more or less smooth 
function in
the whole range, whose asymptotics are given by \eq{c8}.

For example, one can calculate the instanton contribution to the total  
width by integrating the
spectrum. We use the general relation of the type (\ref{c3}) or
(\ref{c5}) to write the total width as minus the integral of the right-
hand side of \eq{spectrum} (without step-functions) from $q_0=-
\infty$ to
$0$. In the leading order in $m_Q \rho$, the integral is easily 
performed
by parts, and is determined near the vicinity of the point $q_0=0$,
immediately yielding, exactly, the total width (\ref{scalwidth}).

It is not difficult to see how this works in the most general case. 
Suppose we
study, for example, the ``semileptonic decays'' $Q\ra q+\ell +\nu$. If 
we can
measure momenta of both $\ell$ and $\nu$, $\;p_{\ell, \nu}$, the 
differential
distribution is merely given by
\beq
d^2\Gamma\; = \; \Phi(p_\ell,p_\nu) \cdot \Im dT(m_Qv-
(p_\ell+p_\nu))\;\;.
\label{c10}
\eeq
Now we want to integrate over the momentum of neutrino and 
determine the
instanton contribution to the charged lepton spectrum. We then keep 
$p_\ell$
fixed and integrate $\Im dT$  with the phase space, $\Phi$, 
depending on $p_\nu$.
If the phase space factor were an analytic function of $p_\nu$, 
the constraints 
(\ref{c3}) or (\ref{c6}) would ensure the vanishing of the integral. 
 
However,
the phase space contains a step-function at $E_\nu=0$, and only for 
this 
reason does one get a nonvanishing integrated width due to 
exponential 
terms.
Then the relation (\ref{c7}) can be used to represent the 
integrated
effect as the effect from the kinematic boundary and above (i.e., over 
the domain
where neutrino ``carries away'' negative energy), which is a hard 
domain where
the expansion is applicable~\footnote{The 
same reasoning equally applies, of course, to the case of the massive
final particles as well. In this case it is convenient to phrase this 
consideration 
in the frame where $m_Qv-p_\ell$ has only a timelike component. 
The phase
space integral is not a polynomial anymore when $m_\nu\ne 0$, but 
is the 
step-function at $E_\nu=m_\nu$ multiplying a fractional power 
function. Then
one merely must put the corresponding phase space factor $P$ in 
\eq{c7}
under the sign of $\Im$ in the right hand side. This does not change 
the
general reasoning presented here.}.
Since in this domain $\Im dT(\kappa^2)$ is a rapidly oscillating
function $\sim \sin{(2\kappa\rho)}$, the integral is determined by 
its 
lower
limit; the expansion in $1/Q$ is obtained by integrating by
parts, and in the leading approximation given by the corresponding
derivative of $\Im dT(\kappa^2)$, from which we must keep only 
the 
derivatives of
$\sin{(2\kappa\rho)}$. The exact coefficients combine to yield just 
the 
value 
of the neutrino propagator at the point which enters in the direct 
calculation
of the lepton spectrum, i.e. $G_\nu(x)$ at $x=2\rho
(m_Qv-p_\ell)/|m_Qv-p_\ell|\,$.

To illustrate the last assertion, we must remember that to get the 
total
width in the leading approximation, we need to integrate the
differential width near $p_\nu=0$ keeping trace of only the 
non-analytic
part due to the neutrino phase space,  and the oscillating part,
$\sin{(2\rho\sqrt{(m_Qv-p_\ell-p\nu)^2})}$, in the differential
distribution. All other constituents of the amplitude can be
approximated by their values at $p_\nu=0$. 

To see that this
integration automatically yields the proper factor, we can use the
following trick: compare this integral with the calculation of the 
neutrino  Green function in the coordinate space by taking the 
Fourier
transform of its momentum representation. Let us take the Fourier
integral by closing the contour of integration over $k_0$ by its
physical residue (we assume here  that the zeroth coordinate 
coincides
with the direction of the vector  $m_Qv-p_\ell$):
\beq
G_\nu(x)\;=\; \int\frac{d^4k}{(2\pi)^4 i} G_\nu(k) \;{\rm e}\,^{ikx}=
\int\frac{d^3k}{(2\pi)^3 2k_0} G_\nu(\vec k) \;{\rm e}\,^{ikx}\;\;.
\label{c11}
\eeq
The large-$x$ asymptotics of the last integral is determined by the 
behavior of the phase space factor $d^3k/k_0$ at small $|k|$ where it
is  non analytic. This factor, on the other hand, is exactly the same
as in the double distribution  if one identifies $k$ in the above
calculation with $p_\nu$ (both are nothing but $d^4k\,
\de_+(k^2-m_\nu^2)/(2\pi^4)$). If  one chooses the coordinate $x$ in
such a way as to have the same  oscillating exponent ${\rm 
e}\,^{ikx}$
on the right hand side of \eq{c11} as in $dT \sim {\rm
e}\,^{2i\rho\sqrt{(m_Qv-p_\ell-p_\nu)^2}}\,$,  identifying $k$ with
$p_\nu\,$,  the values of the integrals will also coincide. The last 
condition just fixes the above stated value of the coordinate. This
matching is rather obvious since the oscillating factor in $dT$ came
originally from  evaluating the exponential ${\rm
e}\,^{i(m_Qv-p_\ell-p_\nu)x}\,$ at the saddle point $x$, which enters
the neutrino  Green function.

Therefore, integrating over the neutrino momentum in the decay, we
recovered  our direct recipe of calculating the lepton  spectrum,
namely  considering the problem as a two-body one, but using $dT$ 
given
by  the product of the quark propagator in the   instanton field, {\em 
and} the  neutrino propagator at the saddle point
$x^2=-4\rho^2$. Since the integrated  effect  comes from the zero
momentum of neutrino, it is governed by the  same exponential
(oscillating) factor determined by $m_Q$ and $p$,  $|\;m_Qv-p_\ell|$.
We can  repeat the very same consideration once more integrating 
now
over the energy of the charged  lepton; the leading term in the
integral again comes from $p_\ell=0$, where the phase  space of the
lepton is non-analytic, the nearby integration reproduces 
$G_\ell(-2i\rho,0)$, the argument of the neutrino  Green function
becomes the same, and we arrive at the total width obtained in the
direct way in  the preceeding subsection.

\section{Instanton Contribution to Heavy Quark Decay Rates -- Real 
QCD}

We now proceed to the case of actual $b$ and $c$ quarks; the 
modification
required here is  accounting  for the quark spins.
In principle, the analysis goes along the same lines  as for the scalar 
quarks.
However, since we deal with  instantons -- 
topologically-nontrivial 
configurations
of the gauge field --
the massless quarks  acquire zero modes. They manifest the
intervention of the infrared, long-distance effects in the presence of
the
instantons. This is an obvious defect of our 
simplified
one-instanton ansatz. If we used the topologically trivial 
(but
nonperturbative) configurations, the problems with the zero modes 
would be
absent.

Technically, this problem emerges already at 
the very
first step:  the Green function of the massless quark is not defined 
in 
the
field of one instanton since the Dirac operator has a zero mode. 
In order to perform estimates similar to the ones described in the 
previous
sections, we need to regularize  the massless quark Green functions 
in 
the 
infrared. We
 do it in the most naive way: introducing a small mass term $m_q$. 
Now, the Green functions are well-defined but they have terms 
which 
behave like
$1/m_q$, which show up wherever the chirality-flip quark 
amplitudes occur. However, 
when the
weak interactions of the quarks are purely left-handed, the problem 
disappears
since the zero modes do not contribute, and we can put 
$m_q=0$ in the
end. We are aware, of course, that this procedure is not fully 
self-consistent, but, hopefully, it works satisfactorily for our limited 
purposes -- revealing 
the correct exponential dependence, as well as the  power of 
$m_Q$, in the
pre-exponent. It seems quite plausible that only the overall 
numerical coefficient will be 
modified in a
more accurate analysis.

Otherwise, the calculations go with  minimal modifications. 
Let us outline the treatment of the semileptonic width.
Once 
again, we
write the transition operator (the integration over $\rho$ will be 
restored at
the very end)
\beq
\hat{T}(r_0)\;=\;
\frac{G_F^2|V_{Qq}|^2}{2}\;
\int \bar{\tilde Q}(x)\Gamma_{\mu} S (x,0) \Gamma_{\nu}
\tilde Q(0) L_{\mu\nu}(x)\;{\rm e}\,^{i(m_Qv-r)x}\,d^4x
\;d^4z
\label{qoper}
\eeq
where the weak polarization tensor for the lepton pair $L_{\mu\nu}$ 
and the
weak vertices are 
\beq
L_{\mu\nu}(x)\;=\;-
\frac{2}{\pi^4}\frac{1}{x^8}(2x_{\mu}x_{\nu} -
x^2\de_{\mu\nu}) \;\;,\;\;\;
\Gamma_\mu = \gamma_\mu(1-\gamma_5) \;\;.
\label{L}
\eeq
The heavy quark fields, $\tilde Q$, are nonrelativistic  as in 
\eq{exp}, and
are originally bispinors. However, solving the Dirac equation of 
motion in
the limit $m_Q\rho \gg 1$ we get the fields in the form of \eq{pexp}. 
The color  matrix $U$ has the same form as in \eq{u}.

The Green function of the light quark in the
instanton background, $S(x,y)$, is expanded in $m_q$ and has the 
following
form \cite{Brown,oneinstR}:
$$
S(x,y) = -\frac{1}{m_q}P_0(x,y) + G(x,y) + m_q\tilde{\Delta}(x,y) +
O(m_q^2)
$$
where $P_0$ is the projector on the zero modes ($P_0$  does not 
contribute to 
\eq{qoper} since it flips chirality.) Therefore, we can merely put 
$m_q=0$ to
arrive at  
$$
\frac{1-\ga_5}{2}S (x,y)\frac{1+\ga_5}{2}
\;=\; \frac{1-\ga_5}{2}G (x,y)\frac{1+\ga_5}{2}\;=
$$
$$
\frac{1-
\ga_5}{2}\left\{
\frac{-\Delta\gamma}{2\pi^2 \Delta^4}(1+ \rho^2/ \zeta^2)^{-1/2}
\Bigl(1+ \rho^2/\eta^2 \Bigr)^{-1/2}
\cdot \Biggl(1 +\frac{\rho^2 (\tau^+ \zeta)(\tau \eta)}{\xi^2
\eta^2}\Biggr)\;+ \right.
$$
\beq
\left.
\;
\frac{-1}{4\pi^2 \Delta^2 \zeta^2 \eta^2}(1+ \rho^2/ \zeta^2)^{-1/2}
(1+ \rho^2/\eta^2)^{-1/2} 
\Biggl( \frac{\rho^2}{\rho^2 + \eta^2} (\tau^+ \zeta)(\tau 
\Delta)(\tau^+
\gamma) (\tau \eta) \Biggr)\right\}
\label{73}
\eeq                                  
where
$$
\zeta=x-z\;\;, \;\;\;\eta=y-z\;,\;\; \Delta=x-y\;\;.
$$
Now,  the calculation differs from the case of the scalar quarks in
minor
technical details, namely,  the different power of $(x^2+\rho^2)$ in 
the
denominator, and the presence of the $\gamma$ matrices in the 
numerator. At the 
saddle point, $U=1\,$, the heavy quark fields enter at the origin, the 
leptonic tensor must be evaluated at
$x_{*}=(2i\rho, \vec 0)$, and the final result is 
\beq
\hat T (r_0)\;= \; 
-\,\frac{G_F^2|V_{Qq}|^2}{4\pi^2\rho^8\left(i(m_Q-r_0)\right)^3}
{\rm e}\,^{ 2i(m_Q-r_0)
\rho} \;\;
\bar{{Q}}(0)\:i\ga_0\,P\: {Q}(0)
\label{tsl}
\eeq
where $P$ is the projector onto the instanton color $SU(2)$ subgroup,
$P=(1,1,0)_{\rm diag}$. Since 
$$
\matel{H_Q}{\bar Q\, \ga_0\, Q}{H_Q}\;=\;2M_{H_Q}
$$
we finally obtain
\beq
\Gamma_{\rm sl}^{\rm I}\;=\; 
-\frac{2}{3}\:
\frac{G_F^2|V_{Qq}|^2}{2\pi^2\rho^8\;m_Q^3}\sin{(2m_Q\rho)}
 \; .
\label{width}
\eeq
In terms of the  free quark semileptonic width, 
$$
\Gamma_0\; =\; \frac{G_F^2m_Q^5|V_{Qq}|^2}{192\pi^3},
$$
we  get
\beq
\Gamma_{\rm sl}^{\rm I}\; = \; - \Gamma_0\:\frac{2}{3}\,
\frac{96\pi}{(m_Q\rho)^8}
\,\sin{(2m_Q\rho)}\;  .
\label{wid}
\eeq

In a similar manner it is easy to find  the expression for the 
differential 
distributions in
the semileptonic decays. We quote here the expression for the 
 instanton-induced lepton spectrum (in its hard part, i.e. far enough 
from the end point) 
in the same approximation,
$$
\frac{m_b}{2\Gamma_0}\,\frac{d\Gamma^{\rm I}(b\ra 
u\ell\nu)}{dE_\ell}\;=\; 
-\frac{2}{3}\:
\frac{48\pi}{(m_b\rho)^5}\; \epsilon^2\,\left(1-
\frac{\epsilon}{2}\right)
\:\frac{\cos{\left(2m_b\rho\sqrt{1-\epsilon}\right)} }{(1-
\epsilon)^{5/2}}
\; ,
$$
\beq
\frac{m_c}{2\Gamma_0}\,\frac{d\Gamma^{\rm I}(c\ra 
s(d)\ell\nu)}{dE_\ell}\;=\;
-\frac{2}{3}\:
\frac{48\pi}{(m_c\rho)^5}\; \epsilon^2(1-\epsilon)^2
\:\frac{\cos{\left(2m_c\rho\sqrt{1-\epsilon}\right)} }{(1-
\epsilon)^{5/2}}
\; .
\label{slspectr}
\eeq
Here $\epsilon=2E_\ell/ m_b$ or $\epsilon=2E_\ell/m_c$ for the 
two decays,
respectively. 
It is immediately seen explicitly, using the technique 
described in
Sect.~7.2, that the integral over the spectrum reproduces the total 
semileptonic
width (\ref{width}).

Now let us proceed to the  nonleptonic decays with the massless 
quarks in the final state. Repeating the derivation above we get 
the non-leptonic width~\footnote{We should note that here the 
subleading
effects are probably significant: even within our simple saddle point
calculation we formally had to discard, e.g. terms of the type 
$9!!/(2Q\rho)^5\simeq
(4/(2Q\rho))^5.$ }
\beq
\Gamma^{\rm I}_{nl}\;=\; 
\frac{2}{3}\:
c_-^2 \:\Gamma_0  
\frac{256\pi}{15 (m_Q\rho)^4}
\,\sin{(2m_Q\rho)}\;  
\label{nlwid}
\eeq
where $c_\pm$ are the standard color factors due to the hard gluons 
in the weak vertex \cite{GLAM}.

For completeness we also give the instanton contribution to the
inclusive radiative rate of the type $b\ra s+\ga$, 
\beq
\Gamma^{\rm I}_{ \ga} \;\simeq\; 
\frac{2}{3}\:
\Gamma^{\rm 0}_{\ga}
\, \frac{12\pi}{(m_Q\rho)^6}
\,\sin{(2m_Q\rho)}\; .
\label{radwid}
\eeq
The above estimate refers to the yield of photons with all energies. 
In experiment one cuts off the low-energy photons, however. 
According to the previous discussion, the introduction of the lower 
cut off can 
 change the estimate of the duality deviations. We will not submerge  
into further details regarding this effect here.  

\section{Numerical estimates} 

In this section, we present  numerical estimates of the possible 
violations
of the local duality in our  model. The 
effects  rapidly decrease  with the 
energy release.  They can be quite noticeable at intermediate
energies, however.  
The inclusive  decay width of the $D$ meson, and
 is 
expected to be one of the prime suspects.  Indeed, with the mass of 
the $c$ 
quark only
slightly over  $1\GeV$, one can expect sizable violations of duality. 
Another case of  potential concern is the 
hadronic width of  $\tau$. This width is also saturated  at a similar 
mass scale.  In these two cases, there exists at least some
(quite incomplete, though) empiric information.  
It is natural to treat one of them as a reference point,
in order to adjust the parameters of the model. Basically, we have 
only one such parameter, the overall normalization of the instanton 
density $d_0$, see Eq. (\ref{deldens}).  We will use the semileptonic 
$D$ 
decays
for this purpose. Then the second problem ($\tau$ decays) can be 
used as a check
that the model is qualitatively reasonable and  does not lead to gross 
inaccuracies. As a matter of fact, this was already demonstrated
in Sect. 5.1. Encouraged by this success we then take the risk to use 
the model 
for numerical estimates of the duality violating effects
in various $B$ decays. Although our model is admittedly imperfect, 
the numbers obtained can hopefully be viewed as  valid  
order-of-magnitude estimates.  

\subsection{$\Gamma_{\rm sl}(D)$}

This decay was numerically analyzed in the heavy quark expansion 
more than once  
\cite{luke}--\cite{upset}, 
with quite 
controversial conclusions. We will summarize here our current point 
of view \cite{BDS}, deferring a brief
discussion of the literature until Sect. 11.

The parton  semileptonic $D$ decay width is given by 
\beq
\Gamma_0(D\ra l\nu X_{s,d})\; =\;
\frac{G_F^2m_c^5}{192\pi^3}\;\simeq\; 1.03\cdot 10^{-13} 
\GeV\;\;\; 
\mbox{ at }\,m_c=1.35 \GeV
\; ,
\label{d1}
\eeq
where   the strange quark mass is neglected. The comparison with 
the experimental value 
\beq
\Gamma_{\rm exp}(D\ra l\nu X)
\;\simeq\; 1.06\cdot 10^{-13} \GeV
\label{d2}
\eeq
seems to be very good. However, there are corrections to the free 
quark 
estimate (\ref{d1}); both the \prt and non\prt corrections calculated 
within
$1/m_c$ expansion work together to noticeably decrease the 
theoretical prediction. 

\vspace{0.3cm}

{\it (i) Perturbative corrections}

\vspace{0.1cm}

The one-loop perturbative correction to the width in the 
four-fermion decay is known since the mid-fifties \cite{QED2}; 
for QCD one gets the factor $\eta_{\rm pert}$ multiplying the 
theoretical formula for the width (i.e. $\Gamma\ra\Gamma_0 
\eta_{\rm pert}$), 
\beq
\eta_{\rm pert}\;=\; 1- \frac{2}{3}\left(\pi^2-
\frac{25}{4}\right)
\frac{\alpha_s}{\pi}  \, .
\label{d3}
\eeq
 This factor obviously decreases the theoretical prediction for the 
width, 
the question is how much. The answer for $\eta_{\rm pert}$ is not
 as obvious as it might
seem, and depends on how $m_c$ is defined.

Equation (\ref{d3}) implies that one 
uses the (one-loop) pole mass of the $c$ quark in the inclusive rate. 
(As well-known, the notion of the pole mass is ill-defined 
theoretically
\cite{pole,gurman,bpole}. 
It is safer to use the Euclidean mass, which also pumps away some of 
the 
$\alpha_s$ corrections from the explicit correction factor 
$\eta_{\rm pert}\,$. 
This decreases the mass and the coefficient in the correction 
simultaneously.  The product $m_c^5 \eta_{\rm pert}$ is numerically
stable, however, and for our limited purposes we can stick to Eq. 
(\ref{d3}) and the one-loop pole mass.) 
The one-loop pole mass was numerically evaluated, say, in the 
charmonium sum 
rules,   yielding \cite{csr}  the number 1.35 GeV quoted above. 
This number is also in a good agreement with the heavy quark 
expansion for the difference $m_b - m_c$ 
\cite{AFMN}, 
\beq
m_b - m_c\; = \;\overline{M}_B \;-\; \overline{M}_D \;+ \;\mu_\pi^2 
\; 
\left(\frac {1}{2m_c} -\frac {1}{2m_b}\right) +{\cal 
O}\left(\frac{1}{m_Q^2}
\right)
\label{d5}
\eeq
where
$$
\overline{M}_{B,D} \;=\;\frac{M_{B,D}+ M_{B^*,D^*}}{4}\;\;.
$$
Here, and in what follows, we use the notations
$$
\mu_G^2\; = \;\frac{\matel{B}{\,\bar b\,\frac{i}{2}
\sigma_{\mu\nu}G^{\mu\nu}\, b\,}{B}} {2M_B}
\approx  \;\frac{3}{4} 
(M_{B^*}^2 -
M_B^2)\, ,
$$
and 
\beq
\mu_\pi^2\; = \;\frac{\matel{B}{\,\bar b  (i\vec{D}\,)^2\, b\,}{B}} 
{2M_B}
\; .
\label{d6}
\eeq
Substituting   $m_b^{\rm pole}\simeq 4.83 \pm .03 
\GeV$ \cite{Vol}, and a reasonable value of $\mu_\pi^2$ (see below),
in Eq. (\ref{d5}), 
we again end up with  $m_c^{\rm pole}\simeq 1.35\GeV$. 
 
Controversial
statements can be found in the literature concerning the   value 
of  $\mu_\pi^2$, (associated mainly with its different understanding)
but the issue appears to be numerically unimportant for our 
purposes.

Now, one has to establish the normalization point of $\alpha_s$ in Eq.
(\ref{d3}). Luke {\em et al.}  suggested \cite{luke2} exploiting the  
BLM prescription \cite{BLM} for this purpose;  it must be 
done in
strict accord with the treatment of the mass. 
This leads to 
\beq
\eta_{\rm pert}\; \approx \; (1-0.25)
\label{d7}
\eeq
(for further details see
\re{upset}).  This number turns out to be stable against the inclusion 
of the ${\cal O}(\alpha_s^2)$, and higher order corrections estimated 
in 
a certain approximation \cite{bbb}. 

\vspace{0.3cm}
{\it (ii) Nonperturbative corrections}

\vspace{0.1cm}

Now let us examine the non\prt $\;$ corrections. There are no 
corrections to the width that scale like $1/m_c$ , and the 
leading ones are given by the 
$1/m_c^2$
terms
\beq
\Gamma_{\rm sl} (D)\;=\;\Gamma_0\eta_{\rm pert}\;\left(1-
\frac{3\mu_G^2}{2m_c^2}-
\frac{\mu_\pi^2}{2m_c^2}\right)\; .
\label{d7a}
\eeq
While $\mu_G^2$ is known, see above, $\mu_\pi^2$ is not yet 
measured in 
experiment. We
have to rely on theoretical  arguments, which, unfortunately, are not 
completely settled yet. 

The original QCD sum rules estimate 
\cite{Ball}
yielded 
\beq
\mu_\pi^2\;=\;(0.5\pm 0.1)\GeV^2\; . 
\label{d8}
\eeq
 We believe that the value $\mu_\pi^2\approx  0.5 \GeV^2$  is 
the most reasonable estimate available at 
present. It  matches a general inequality,
\cite{motion,Vol2,vcb}
\beq
\mu_\pi^2\; > \; \mu_G^2\; \simeq 0.36\GeV^2\; ,
\label{d9}
\eeq
and a more phenomenological 
estimate of Ref. \cite{third}.  Moreover, Eq. (\ref{d8}) is  marginally
consistent with the first attempt of extracting $\mu_\pi^2$ directly 
from 
data \cite{Gremm}, keeping in mind the theoretical uncertainties
encountered there (other analyses are in progress now). 
 In any case, the effect of the kinetic operator on the semileptonic  
$D$ width 
is modest, so that the impact of the uncertainties debated in the 
literature  can lead to at most  $\sim 5 \%$ 
change in $\Gamma_{\rm sl}(D)$.\vspace*{0.4cm}

Assembling all pieces together, numerically we get
\beq
\Gamma^{\rm sl}_{\rm th}(D)\; \simeq\;\Gamma_0\;\left(1-0.25-
0.3-0.15\right)
\label{d10}
\eeq
where the corrections  in the parentheses stand for  the \prt 
correction,
the chromomagnetic and the kinetic energy terms, respectively. 
Thus, one is left with less than a half of the experimental width. 
According 
to
\re{BDS}, the next order non\prt ${\cal O}(1/m_c^3)$ effects 
apparently do not
cure -- and possibly deepen -- the discrepancy.

Pushing the numerical values of the 
parameters above, within uncertainties, to their extremes (but still, 
within acceptable limits), one can somewhat narrow the gap, but it is 
certainly impossible to eliminate it completely. Therefore, it is 
natural to conclude that the observed discrepancy  in $\Gamma_{\rm 
sl}(D)$,
at the level of several dozen percent, is due to duality violations. 

We will make this bold assumption. 
The instanton contribution to $\Gamma_{\rm sl}(D)$, Eq. (\ref{wid}),
is then convoluted with the instanton density (\ref{deldens})
to yield
\beq
\Gamma_{\rm sl}^{\rm I} (D)\; =\;-\Gamma_0 
\:\frac{2}{3}\,
d_0\, 
\frac{96\pi}{(m_c\rho_0)^8}
\,\sin{(2m_c\rho_0)}\;  .
\label{widconv}
\eeq
Ignoring the sine on the right-hand side \footnote{The value
of $\sin{(2m_c\rho_0)}$ is sensitive to how close the argument is to 
$k\pi$. This proximity is a very model-dependent feature, sensitive 
to small variations of parameters. Since we are aimed at conservative 
estimates all sine factors here and below will be consistently
put equal to unity.},
and requiring this contribution to be
$0.5 \,\Gamma_0$, we obtain  
\beq
d_0 \approx 9\times 10^{-2}
\label{estd}
\eeq
(the values $\rho = 1.15 \, {\rm GeV}^{-1}$ and $m_c = 1.35 \, {\rm 
GeV}$ are used).  We will consistently exploit the above  
values of
$d_0$ and $\rho_0$  in all numerical estimates in Sect. 9.2. 

If our approach is applied to the hadronic $\tau$ width,
deviation from duality comes out to be  
\beq
\frac{\Gamma^{\rm I}(\tau\ra {\rm hadrons})}{\Gamma_0\;(\tau\ra 
{\rm hadrons})}\sim
 \, d_0\,\frac{4\pi^2}{(m_{\tau}\rho_0)^6}  \approx 5\times 10^{-2} ,
\label{tauwid}
\eeq
i.e. quite reasonable. The fact that the numbers come out 
qualitatively reasonable in this case is also demonstrated by 
Fig. \ref{fig1}. Let us note in passing that the $5\%$ uncertainty in 
$\Gamma (\tau\ra {\rm hadrons})$ translates into $\sim 30\%$ 
uncertainty in
the value of $\alpha_s (m_\tau )$.

\subsection{Duality violation in $b$ decays}

What is the magnitude of the 
anticipated 
effects in other situations where they are not yet determined 
experimentally,
say, in beauty decays? Within our model the answer can be given.
In what follows we will use expressions obtained in Sect. 8, 
convoluted with the instanton density (\ref{deldens}), ignoring sines 
and cosines in numerical estimates. In this way we expect to get an 
upper bound on the duality violating contributions. This expectation 
is based on the following: (i) smearing with a more realistic 
finite-width instanton density will inevitably result in some extra 
(exponential) suppression, compared to the delta-function density;
if the instanton density is distributed rather narrowly, the above 
suppression
plays no role in $D$ and $\tau$ but will presumably show up
at  high energy releases characteristic to $B$ decays;  
(ii)~replacing sines and cosines by unity we increase
the estimated value of the duality violations. 

Let us first address the simplest case, $b\ra u\, \ell \nu$ 
semileptonic
width, where the energy release is the largest. To abstract as much 
as 
possible
from the untrustworthy details of the instanton model, we can 
merely use the
scaling behavior \eq{wid}. In other words,
in all expressions below we keep only the pre-exponential factors.
 Then  
\beq
\frac{|\Gamma^{\rm I}(b\ra u\, \ell\nu)|}{\Gamma_0 (b\ra u\, 
\ell\nu)} 
\; \sim \;\frac{2}{3}\,
d_0
\;   
 \frac{96\pi}{(m_b\rho_0)^8}
 \; \;\approx  
2\times 10^{- 5}   \; . 
\label{f3}
\eeq

A similar estimate for the radiative transition, $b\ra 
s+\ga$,
 based on \eq{radwid}, yields
\beq
\frac{|\Gamma^{\rm I}(b\ra s+\ga)|}{\Gamma_0(b\ra s+\ga )} 
\sim \;
\frac{2}{3}\,
d_0\,
\frac{12\pi}{(m_b\rho_0)^6}
 \;
\approx \;  8\times 10^{-5}\;\;.
\label{f4}
\eeq

Next, we move on to processes with a heavy quark in the final 
state. Of particular  
interest  are duality violating  effects in the Kobayashi-Maskawa 
allowed $b
\ra c$ transitions. Consider first the semileptonic decays
$B\ra X_c\,\ell \nu$. As was discussed in Sect. 6.2, the instanton 
result  for this process vanishes in the leading saddle-point 
approximation, while the subleading terms near the saddle point 
have not been calculated. To get an upper bound on the duality 
violations 
in the $b\ra c$ transition, in a rough approximation, we neglect all 
these subtleties and
merely use our expression for the $b\ra u$  replacing $m_b$ 
by $m_b-m_c$,
\beq
\frac{|\Gamma^{\rm  I}(B\ra c\,\ell \nu)|}{\Gamma_0(B\ra c\,\ell 
\nu )}
\; \sim \;\frac{2}{3}\,
d_0 \,
\frac{96\pi}{\rho^8_0 (m_b-m_c)^8 }
  \; \;\approx 3\times  10^{ -4}\; .
\label{f8}
\eeq

Summarizing,  the duality violating  corrections are expected to  be 
negligible 
in the
semileptonic and radiative $B$ decays, even in the $b\ra c$ 
transitions. 

Let us proceed now to nonleptonic decays. Here, according to our 
model, the 
situation may somewhat change: deviations from duality jump up.  
Intuitively it is clear that the smallest effect is expected in the 
channel $b\ra u\bar u 
d$ where the energy release is the largest. Specifically, 
\beq
\frac{|\Gamma^{\rm I} (b\ra u\bar u 
d)|}{\Gamma_0 (b\ra
u\bar u d)}\;\sim  \; 
\frac{2}{3}\,
d_0\: c_-^2\:
\frac{256\pi}{45 (m_b\,\rho_0)^4}
 \, \; \approx   10^{-3}\; ,
\label{f11}
\eeq
where 
$$ 
\Gamma_0 (b\ra u\bar u d)\;=\; 3 \, \Gamma_0 \; =
\; \frac{G_F^2\,m_b^5\;|V_{ub}|^2}{64\pi^3} \,
$$
 and
$$ 
c_-=c^{-2}_+=\{\alpha_s(m_b)/\alpha_s(M_W)\}^{12/23}\approx 1.3 .
$$

\vspace{0.2cm}

The effect further increases in the case of  Kobayashi-Maskawa 
allowed nonleptonic transitions
$b\ra c\bar u d$.  

  Considering the
$c$ quark as a static heavy quark which, according to section 6.2,
does not interact with the instanton field at the saddle point, we 
obtain the following expression:
$$
\frac{|\Gamma^{\rm I} (b\ra c\bar u d)|}{\Gamma_0
(b\ra c\bar u d)}\; \sim \;2d_0\; \frac{7c_+^2+3c_-^2 +2c_+c_-}
{12}
\;\times
$$
\beq
 \;\frac{16\,\pi^{5/2}\:(m_b-m_c)^{1/2}}{m_b^5\;\rho_0^{9/2}}
\left(\frac{m_c}{m_b}\right)^{3/2}
\; \approx 3\times 10^{-3}\, .
\label{ura}
\eeq 
Here
$$
\Gamma_0 (b\ra c\bar u d)\;\simeq\; 3 \; \Gamma_0  
 \cdot 0.5 \;=
\; 0.5\,\frac{G_F^2\;|V_{cb}|^2\;m_b^5}{64\pi^3}  ,
$$
with the factor $0.5$  due to the  kinematical suppression in the 
phase space associated with the $c$ quark 
mass. The peculiar expression with the $c_\pm$ factors above
 emerges from  the color matrices in the weak Lagrangian 
and in the quark Green functions after averaging over orientations
of the $SU(2)$ instanton over the color $SU(3)$ group. 

However, one may worry that
the $c$ quark is not heavy enough since  the parameter
$m_c\rho \sim 2 $ is rather close to unity. Then, we may also try to 
consider 
another limiting case, and treat the $c$ quark as massless, keeping
$m_c$ only in the energy release. Then 
applying the technique from Sect.~6.1, we obtain  
\beq
\frac{|\Gamma^{\rm I}(b\ra c\bar u d)|}{\Gamma_0
(b\ra c\bar u d)}\;\sim \;2\;
\frac{2}{3}\,
d_0\;c_-^2\,
\frac{256\pi}{45 \rho_0^4 (m_b-m_c)^4}
 \; \;
\;\approx  8 \times 10^{-3}
\; ,
\label{f12}
\eeq
which is close, numerically, to the previous estimate. 
Note, the coefficients in front of
$c_+$ and $c_-$ are different since the
heavy quark does not have  color structure in the
propagator while the massless quark does.

The instanton contribution is enhanced by an order of magnitude due 
to the fact   that the energy release is by a factor,
 $(m_b-m_c)/m_b \approx 0.7 $, smaller
then in the channel with massless quarks, and due to the 
kinematical suppresion in the partonic width.   

Finally, let us discuss the 
 duality violating contributions in the transition with two heavy 
quarks in the final state,  $b \ra 
c\bar c s$, where they are believed to be the largest, 
for an obvious reason: the energy release is the 
smallest
\footnote{Phenomenological analyses of the $b \ra 
c\bar c s$ channel are presented, e.g. in recent works  \cite{Duni}.}.
This natural expectation does not contradict
our model, although the enhancement we literally get is rather modest, 
cf. Eqs. (\ref{ura}) and (\ref{f15}). The 
SV approximation for the $c$ quarks in the transition $b \ra 
c\bar c s$
is justified; the simplest saddle point 
evaluation yields a non-vanishing effect due to the presence of the
$s$ quark in the final state. Performing the saddle point evaluation,
one obtains
\beq
\Gamma^{\rm I} (b\ra c\bar c s\,(d)) \;\sim \;d_0\; 
\frac{2c_+^2 +c_-^2}{3}\,
\frac{ G_F^2|V_{cb}|^2\;|V_{cs}|^2\, m_c^3}{\pi m_b^{3/2}(m_b-
2m_c)^{3/2}\rho_0^5}\, 
\sin{(2\rho_0(m_b-2m_c))}\, ,  
\label{f13}
\eeq
with  
$$ 
\Gamma_0 (b\ra c\bar c s)\; \simeq  \;3 \; \Gamma_0 \cdot 
0.15 =
\; 0.15\,\frac{G^2\;|V_{cb}|^2\;|V_{cs}|^2\;m_b^5}{64\pi^3}   ,
$$
and 0.15 coming from phase space suppression due to two $c$ 
quarks in the final state. 
Numerically, 
\beq
\frac{|\Gamma^{\rm I} (b\ra c\bar c s)|}{\Gamma_0
(b\ra c\bar c s)} \;\sim 7\;d_0 \; 
\frac{2c_+^2 +c_-^2} {3} 
\frac{64\;\pi^2}{(m_b\rho_0)^5}
\frac{m_c^3}{(m_b(m_b-2m_c))^{3/2} }
\approx  6\times 10^{-3} \, .
\label{f15}
\eeq
 
\vspace{0.2cm}

Concluding this section, let us reiterate our numerical findings. 
Using the saddle point
approximation, and the instanton model to determine the nature of 
the
singularities in the quark propagators, in the complex (Euclidean) 
plane, we are able to derive some scaling 
relations
for the duality violating effects induced by ``soft background" fields. 
Keeping track of only the powers of relevant energies in the 
pre-exponent and   neglecting the rest, 
we observe the following hierarchy:
$$
\frac{\Delta\Gamma^{\rm I}}{\Gamma_0} (b\ra c\bar c s) \;
\sim \;2 \frac{\Delta\Gamma^{\rm I}}{\Gamma_0} (b\ra c u d)\;
\sim \;5 \frac{\Delta\Gamma^{\rm I}}{\Gamma_0} (b\ra u \bar u d)
$$
$$
\sim\; 16 \frac{\Delta\Gamma^{\rm I}}{\Gamma_0} (b\ra c l \nu)\;
\sim \; 75 \frac{\Delta\Gamma^{\rm I}}{\Gamma_0} (b\ra s \gamma)
\;\sim\; 300 \frac{\Delta\Gamma^{\rm I}}{\Gamma_0} (b\ra u l \nu)\, .
$$
 
Our numerical estimates are expected to be upper bounds
within the particular mechanism of duality violations considered in 
the present paper. We hasten to add, though, that there exist 
physically 
distinct mechanisms, e.g. due to hard non-perturbative fluctuations, 
or those which may be somehow related to  the spectator light 
quarks in the initial state, and so on. They deserve a special 
investigation. 

\section{Drawbacks and Deficiencies of the Model}

Our  instanton model of duality violations has obvious shortcomings.
Although qualitatively it correctly captures the essence of the 
phenomenon we want to model -- transmission of a large external 
momentum through
a soft background gluon field -- the one-instanton ansatz itself is too 
rigid to be fully  realistic. It has virtually one free parameter, the 
instanton size, and this is obviously not enough for perfectly
successful phenomenology.  The value of $\rho$ we use 
is even somewhat  smaller than that usually accepted in the  
instanton liquid model \cite{Dodik}. Correspondingly, the oscillation 
period comes out too large. For instance, in    $R(s)$, \eq{b3}, the 
oscillation period is almost four times  as 
large as one observes
experimentally. Figure \ref{fig2} suggests that  the typical oscillation length
 in $R(s)_{\rm exp}$ is $\sqrt s \sim 0.6 \, {\rm GeV}$
    while from Fig.\ref{fig1} 
we get   $\sqrt s \sim\;2.7 \;GeV$. Even if we used the instanton 
liquid 
value,  this would not narrow  the discrepancy in any 
significant way.
Moreover, the data seems to suggest that the oscillation length 
slowly varies as we move to higher energies. Our one-instanton
ansatz is certainly incapable of reproducing this feature.
The lesson we learn is that the soft background field has to be larger 
in scale and more sophisticated in shape. 

Another manifestation of the unwanted rigidity of the one-instanton 
background is the occurrence of  zero modes for massless quarks.
This phenomenon also leads to some inconsistencies in our treatment.
For instance, the two-point function of the axial currents
 would not possess the necessary transversality properties. 
We essentially ignored this problem, keeping in mind the emphasis 
we place on qualitative aspects of duality violations. After all, our 
model is semi-quantitative, at best.

The soft gluon fluctuations crucial in the duality violations are
definitely not the ones constituting the dominant component of
the vacuum. Indeed, if we used the instanton weight
(\ref{deldens}), with $d_0$ fitted to reproduce the  duality violations 
in the $D$  meson semileptonic decay, as the instanton density in the 
liquid model we 
would render this model disastrous. With our density in the 
instanton liquid model we would get the value of the gluon 
condensate $\sim 25$ times larger than it actually is. 

In addition to the above negative features, the model obviously 
misses other mechanisms which might also lead to violations of 
duality.  The most noticeable is the absence of the impact of the 
initial light quarks in $b$ decays. At the very least, one could suspect 
that
they play a role in the formation of the soft gluon medium
which, after the decay, has to transmit a large energy release.
The influence of the initial light quarks would make the duality 
violating 
effects non-universal (they will be different, say, in mesons and 
baryons). At the moment we have no idea how to take 
into account this effect, nor we have any idea of how essential it 
might be numerically. 

On the positive side, we would like to stress again, that the model is 
general enough. 
One considers instantons only as a source of  finite distance 
singularities in the quark Green functions, and for that purpose
they may serve satisfactorily. 
Our procedure
has very little to do with the full-scale  instanton calculations
of the type presented in Refs. \cite{balbb} --
\cite{oneinstR}. In this sense, our calculations are much less 
vulnerable than the standard instanton exercises. The 
finite-distance singularities merely represent the  mechanism
of transmitting a large momentum  through a large number of soft 
``lines", with no hard lines
involved (so that this mechanism  does not appear in the practical OPE).
Our point of view is  pragmatic:  experimental data 
clearly indicate 
 duality violations, with an oscillating pattern, and so we reproduce 
this physical effect 
through fixed-size instantons.  Eventually, comparison with data will 
lead  to a better understanding of
the relevant gluon field configurations and emergence of a model 
free from the drawbacks summarized above. 

\section{Comments on the Literature}

The  present work intertwines many aspects of QCD -- purely 
theoretical and phenomenological -- in one junction; some of these 
aspects are quite controversial and cause heated debates. 
Therefore, it is in order to briefly review the literature
in which  the relevant issues were discussed previously. 

We have already mentioned previous instanton calculations  in 
$R_{\rm e^+e^-}$ and $\tau$ decays
\cite{oneinstR, bal,balbb,nason,?}. Technically, they are very 
instructive and advanced. There are hardly any 
doubts, however, that the solitary instantons considered in these 
exercises
do not represent, in the dynamical sense,  typical relevant vacuum 
fluctuations. The fact that the corresponding estimates  fell short of
the experimentally observed effects is neither  surprising nor 
frustrating. At the same time, the  provocative suggestion of Ref. 
\cite{DS2} to use instantons for abstracting finite-distance 
singularities
in the quark Green functions,  was largely ignored. 
We try to develop this idea to its logical limits. 

Recently, the impact of the small-size instantons was analyzed in the 
spectral distributions of the inclusive heavy quark decays,
within the formalism of HQET \cite{Chy, AFalk}. The ``part larger 
than the whole" paradox was first detected in these works:
the instanton contribution to the spectra 
was found to be parametrically larger than the very same 
contribution to the decay rate.  (The result was divergent
at the boundaries of the phase space. This divergence is due to the 
fact that the isolated instanton density badly diverges at large 
$\rho$, and  the instanton size is regulated by the external energy 
release, which vanishes at the end points.) The solution of this 
paradox was discussed at length above.  
The spectra near the boundaries of the phase space can not be 
calculated point-by-point in the 
present-day QCD. Still, the integrals over the spectra over a finite 
energy range touching the end-point are calculable.  Integrating
the instanton contribution, taking into 
account its peculiar  analytical 
properties, automatically  yields the effect which is determined by 
the far side of the
smearing interval, rather than by the end point domain. The main 
subtlety lies in  the process of separation of a ``genuine" 
instanton contribution
from the regular OPE.  Our procedure automatically avoids double 
counting, an obvious virtue which is  hard to achieve
otherwise. 

The question of whether or not the semileptonic $D$ decays are 
subject 
to noticeable duality violations is more controversial and is debated 
in the literature. Sometimes it is claimed that
the OPE result  (\ref{d7a}) is compatible with 
experimental data, with no  additional terms. The price 
paid is rather high, however: the mass of the $c$ 
 quark is then  pushed up beyond 1.55 GeV  (the value of 
$\mu_\pi^2$ is pushed down almost to zero). 
If the value of $m_c$ 
was that high, one would be in trouble in many other problems, e.g.
the charmonium sum rules \cite{csr}, the
analysis of the $b \bar b$ threshold region \cite{Vol}, and so on.
Moreover, using  a calculational scheme, which relies on large $m_c$,
leads to poor control of the perturbative series -- a fact noted in Ref. 
 \cite{luke2}.
We believe that the value of the product $m_c^5\eta_{\rm pert}$ 
used above is realistic, which inevitably entails violations of 
duality in the $D$ meson semileptonic decays in the ballpark of 
several units times $10^{-1}$. Let us emphasize that our standpoint is  
testable experimentally. One of possible tests is analyzing, say, the 
average lepton energy in the $D$ meson semileptonic decays. 
This quantity is much less sensitive to the value of $m_c$ than the 
total width.
Therefore, the task of detecting deviations from the OPE becomes much 
easier. An indirect proof may be provided by confirmation of  
duality violations in the $\tau$ lepton rate in the ballpark of several 
units times $10^{-2}$.

\section{Conclusions and outlook}

At high energies the inclusive decay rates (e.g. $\tau\ra\nu +$ 
hadrons, or $B\ra X_u +\ell\nu$, or nonleptonic $B$ decays)  are 
represented by the 
sum of the transition 
probabilities into a very large number of  possible final states. 
It looks like a miracle that these complicated sums, with various
threshold factors, final state interactions and so on, reproduce a 
smooth quark (gluon) curve. This {\it duality} is explained by QCD.  If 
the quark (gluon) cross section is calculated by virtue of the 
procedure known as the {\it practical OPE}, one expects that the inclusive  
hadronic cross section coincides with the quark (gluon) curve  at 
large energy releases up to terms which are exponential in the 
Euclidean domain,
and have a very peculiar oscillating pattern in the Minkowski 
domain, where they fall off relatively slowly, at least in the 
oscillation 
zone (Sect. 5.2).  Physically these {\it exponential} terms are 
associated with the
transmission of  large external momenta through the soft gluon 
medium. 

To model this mechanism mentioned above  we suggest 
instanton-motivated estimates. The instantons are used  to abstract  
general features of
the phenomenon and, to some extent, to gauge our expectations. 
We associate duality violations with
the finite distance singularities in the quark Green functions due to
soft background  field configurations. 
The instanton-induced finite distance singularities
produce a pattern of duality violations which closely resembles the 
indications (rather scarce, though) provided by current experimental 
data on $\rm e^+e^-$ and $\tau$ decays.  The free parameters of the 
model are calibrated using these data.

\begin{figure}
  \epsfxsize=12cm
  \centerline{\epsfbox{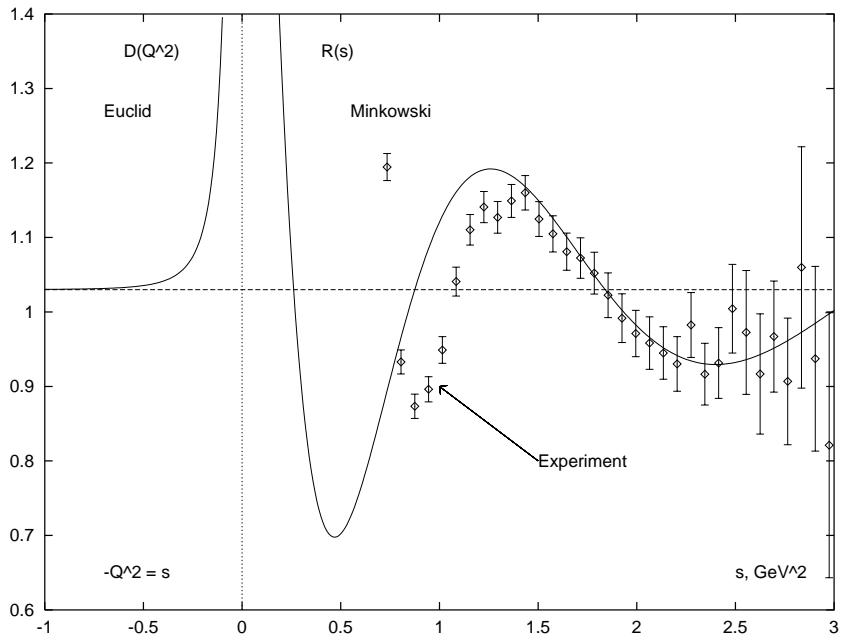}}
  \caption{Experimental data for $R_{V-A}(s)$; the solid line at $s>0$ is
    a Bessel function (see text) drawn to guide the eye through the data
    points. The curve at the Euclidean  $Q^2$ is the corresponding 
    McDonald function. $\;\:D(Q^2)= \frac{1}{\pi}(Q^2d/dQ^2 )\Pi(Q^2)$.  
}
  \label{fig6}
\end{figure}

Let us examine, for instance, Fig. \ref{fig2}. 
which presents  the differential hadronic mass distribution in
$\tau$ decays. From the first glance it is clear that    significant (up 
to $\sim 20\div 30\%$) 
violations of
local duality are present in the whole accessible range. As a matter of 
fact,  the ``oscillating'',
duality-violating part of the effective $V\!-\!A\,\times \,V\!-\!A$
cross section,
 $\de R_{V-A}$, is well approximated, in the region
$0.8\GeV^2-3\GeV^2$, by the function
\beq
\de R_{V-A}\;\simeq\;5\left(\frac{J_1(7.5\sqrt{s})}{7.5\sqrt{s}}-
\frac{2}{7.5^2} \de(s)
\right)\;\;\;\;  (\;s \mbox{ in } \GeV^2\;)\;  .
\label{f1}
\eeq
The right-hand side is about $-0.1$ at $s=2.5\GeV^2$. 
If we take this function literally at all $s$, and 
reconstruct the corresponding correlator $\de\Pi$ in the 
 complex plane, we will find that it  
 has no  $1/Q^2$
expansion at all and, hence, would be omitted in any calculation 
based on the practical OPE.  In
the Euclidean domain, the corresponding $\Pi (Q^2)$  decreases 
exponentially,
\beq
-\,\de \Pi(Q^2)\;\simeq\; 10\:
\frac{K_1(7.5\sqrt{Q^2})}{7.5\sqrt{Q^2}}
\;\; .
\label{f2}
\eeq
We plot $R(s)$ corresponding to \eqs{f1} and \ref{f2} in Fig.~\ref{fig6}.
The value of $\de \Pi(Q^2)$ is 
extraordinarily small, $\sim 3\times 10^{-4}$,  already at
$Q^2=1\GeV^2$. It 
reaches
a ``noticeable" $30\%$ level only at $Q^2=0.1\GeV^2$, at the mass of 
the two pion threshold. And it is still (at least) as important as the 
usual
perturbative corrections in the Minkowski domain, in the physical 
cross section,  at as large of values of $s$ as  $s\simeq 2.5\GeV^2$! 
With 
some reservations we can say that in 
the $\rm e^+e^-$ annihilation and in the
$\tau$ decays we already have direct experimental evidence that 
such effects are significant. 

Less direct -- but still quite convincing -- arguments show that
semileptonic  charm decays also exhibit a similar phenomenon. There 
are good reasons to believe that such a situation is not 
exceptional. We argued that these effects represent a typical
behavior of the  strongly confining interacting theory in  
Minkowski space.                                                   

Accepting our model, with all its drawbacks, for qualitative 
orientation we were able to achieve  
certain  
progress
in relating various pieces of phenomenology to each other.  First, 
we found that the observed $20$ to $30\%$
deviations in  $\rm e^+e^-$ annihilation and in  the spectra of the 
$\tau$ decays are
consistent with  significant corrections to the inclusive
semileptonic $D$ decay rate. At the same time, our
estimates of duality violations rapidly 
decrease with increase of the energy scale, and produce seemingly 
negligible effects in the inclusive decays of beauty.

The strongest duality violations are expected to occur
in the non-leptonic decays of $B$ mesons, especially in those which 
contain two charmed quarks in the final state.
In the transitions $b\ra \bar{c}cs$ they are of order of $1\%$, while   
in the semileptonic and radiative decay rates deviations from duality 
fall off in magnitude to several units $\times 10^{-4}$. 

Although our estimates  are universal -- they do not distinguish, say,
between $B$ mesons and $\Lambda_b$ baryons -- the model {\it per 
se}, taken seriously  actually carries seeds of   
``spectator-dependency''. Indeed, the presence of extra light quarks
in the initial state (baryons of the type $\Lambda_b$) can help lift
the chiral suppression of instantons, enhancing their weight 
compared to the meson case. 
In the case of
$\Lambda_b$ the spectator quarks can naturally saturate the 
instanton
zero modes for $u$ and $d$ quarks  in the diagram incorporating 
both
``Pauli Interference" and ``Weak Scattering"-type processes.
Then it is natural to expect stronger violations of duality.
The argument is quite speculative, of course. 
This issue has  not been investigated in detail. A dedicated analysis
is clearly in order. 

The instanton model we suggest for estimating duality violations 
relies only on the most general features of instanton calculus, 
deliberately 
leaving aside concrete details. Instantons  are taken merely as 
representatives of a strong coherent 
 field configuration which  have fixed size $\rho
\gg 1/Q$, providing the quark Green functions with finite-distance 
singularities.  In this situation we get a transparent picture of the
corresponding duality violating phenomena. Technically, in 
 the  {\it Minkowski 
kinematics} the effect of the finite-$x$ singularities can be viewed as
an additional emission of a spurious particle, a ``ghost", with an 
arbitrary
mass $\kappa$. The mass of the fictitious ghost has a smooth  
distribution,  decreasing as
some power of $1/\kappa$, but oscillating in a more or less universal 
way. Equation (\ref{f1}),
with $7.5\sqrt{s}$ replaced by $2\rho\kappa$, is an example. The 
peculiarity of the $\kappa$ distribution   is a remarkable fact 
that the overall decay rate, with emission of the ``ghost",  is always 
saturated
at the maximal invariant mass of the ghost available in the process
at hand.
This qualitative picture can actually be converted into a kind of 
special diagrammatic technique for the ghost propagation. 
The instanton-motivated estimate of the
strength of the ``ghost coupling'' is probably too crude. However, we 
think that such an approach, in a
generalized form, may prove to be useful in describing violations of 
duality.

Exploring duality violations is a notoriously difficult task. This field 
practically remains {\it terra incognito},  over the
 two decades since the 
advent of  QCD.  Our present attempt is only the first step.
The model itself has an obvious potential for improvement.
Getting rid of the rigidity of
the one-instanton ansatz, one may hope to achieve phenomenological 
success in describing fine structure of the duality violating effects:
the length of the oscillations and its modulations, and so on. 
(The pattern of experimental data suggests, perhaps, the 
presence of more than one scale of oscillations in the duality 
violating component. More accurate data that could provide 
more definite guidelines, are still absent.)

To this end, it is necessary to consider as a background
field a more generic configuration, with more free parameters, say,
an  instanton molecule, or liquid type configurations. It is clear that 
the configurations relevant to the phenomenon under consideration 
\footnote{Remember, these configurations  definitely 
have very little to do with those determining the most essential 
features of the QCD vacuum.} 
have at least two scales built in, and one of them is significantly 
lower than 1 GeV.  Such a  project will require a lot of numerical 
work, however -- an element we wanted to avoid at the first stage. 

Two other promising  directions for explorations of  duality violations 
are two-dimensional models and weakly coupled QCD in the Higgs 
phase. Both directions are much simpler than the actual QCD, and still 
the phenomenon is complicated enough so that the answer is not 
immediately clear. The first attempt of using the 't Hooft model for 
this purpose was made in Ref. \cite{Zhit}, which contains some initial 
observations. The potential of the model is clearly far from being 
exhausted. As for gauge theories with the spontaneous breaking 
of symmetry, calculation of the inclusive two-particle scattering  
near the sphaleron mass,  revealing the typical pattern of the 
cross section, would be extremely instructive for QCD proper
\cite{MS}. 

Let us  note in this respect that  lattice QCD, unfortunately,  can add 
very little, if at all, to the solution of the problem of duality 
violations.  The reason is quite obvious: all lattice simulations are 
done in the Euclidean domain, where all ``exponential" terms are 
indeed exponentially small. Physically interesting and numerically 
important are these effects at large energies, deeply inside the 
Minkowski domain, and very far from
the Euclidean domain where the lattice simulations are formulated. 

\vspace*{.3cm}

{\bf ACKNOWLEDGMENTS:} \hspace{0.2cm} The authors are grateful 
to A.~Vainshtein and M.~Voloshin for useful discussions, and to S. 
Sanghera
for assistance with the CLEO experimental data on $\tau$ decays. N.U. 
also 
benefited from numerous discussions with Yu.~Dokshitzer.
This work was supported in part by DOE under the grant
number DE-FG02-94ER40823.

\end{document}